\documentclass[preprint,superscriptaddress,tightenlines,eqsecnum,
floats,aps,nofootinbib,prd,showpacs]{revtex4}

\usepackage{amsmath,amssymb,amsfonts}

\def\be{\begin{equation}}
\def\ee{\end{equation}}
\def\ba{\begin{eqnarray}}
\def\ea{\end{eqnarray}}
\def\nn{\nonumber}

\newcommand{\hilbert}{\mathcal{H}} 
\newcommand{\abs}[1]{{\left|{#1}\right|}} 
\newcommand{\inner}[2]{{\langle {#1}\vert {#2} \rangle}} 
\newcommand{\ket}[1]{\vert{#1}\rangle} 
\newcommand{\mean}[1]{{\langle {#1} \rangle}} 
\newcommand{\Mean}[1]{{\left\langle {#1} \right\rangle}} 

\newcommand{\Pl}{\ell_\mathrm{Pl}} 
\newcommand{\mubar}{{\bar \mu}} 
\newcommand{\calK}{\mathcal{K}}
\newcommand{\bK}{{\bar{K}}}
\newcommand{\bS}{{\bar{S}}}
\newcommand{\bG}{{\bar{G}}}

\newcommand{\secref}[1]{Sec.~\ref{#1}}

\newcommand{\eqnref}[1]{(\ref{#1})}

\newcommand{\appref}[1]{Appendix~\ref{#1}}
\newcommand{\footref}[1]{Footnote~\ref{#1}}

\begin{document}

\preprint{IGC-08/11-7}

\title{Effective Equations of Motion for Constrained Quantum Systems: A Study of the Bianchi I Loop Quantum Cosmology}

\author{Dah-Wei Chiou}
\email{chiou@gravity.psu.edu}
\affiliation{
Institute for Gravitation and the Cosmos,
Physics Department, The Pennsylvania State University,
University Park, PA 16802, USA}
\affiliation{
Department of Physics, Beijing Normal University,
Beijing 100875, China}

\begin{abstract}
A new mathematical framework is formulated to derive the effective equations of motion for the constrained quantum system which possesses an internal clock. In the realm close to classical behavior, the quantum evolution is approximated by a finite system of coupled but ordinary differential equations adhered to the weakly imposed Hamiltonian constraint. For the simplified version of loop quantum cosmology in the Bianchi I model with a free massless scalar filed, the resulting effective equations of motion affirm the bouncing scenario predicted by the previous studies: The big bang singularity is resolved and replaced by the big bounces, which take place up to three times, once in each diagonal direction, whenever the directional density approaches the critical value in the regime of Planckian density. It is also revealed that back-reaction arises from the quantum corrections and modifies the precise value of the directional density at the bouncing epoch. Additionally, as an example of symmetry reduction, we study isotropy emerging from the anisotropic Bianchi I model in the context of effective equations of motion.
\end{abstract}


\maketitle

\tableofcontents



\section{Introduction}\label{sec:introduction}
The comprehensive formulation for loop quantum cosmology (LQC) in the $k=0$ Friedmann-Robertson-Walker (FRW) model (i.e. spatially flat and isotropic) with a free massless scalar field has been rigourously investigated \cite{Ashtekar:2006rx,Ashtekar:2006uz}, revealing that the big bang singularity is resolved and replaced by the big bounce. Later in \cite{Ashtekar:2006wn}, based on the same principles but implemented with a more sophisticated quantization scheme, a new Hamiltonian constraint was constructed and the big bounce is shown to take place precisely when the matter density enters the Planck regime. These attractive results affirm the assertion that the quantum geometry of loop quantum gravity (LQG) holds a key to avert the breakdown of the classical general relativity and since then have inspired a lot of research on the simplified (but exactly solvable) model \cite{Ashtekar:2007em,Corichi:2007am} and other extensions such as the $k=\pm 1$ FRW models \cite{Ashtekar:2006es,Vandersloot:2006ws}. To further extend the domain of validity, the framework of \cite{Ashtekar:2006rx,Ashtekar:2006uz,Ashtekar:2006wn} was also formulated for the Bianchi I model to include anisotropy \cite{Chiou:2006qq}.

In the Bianchi I model, the similar prediction for the occurrence of big bounces has been anticipated in the analytic investigation of \cite{Chiou:2006qq} and later verified in the numerical computation of \cite{Szulc:2008ar} (for a slightly simplified version of \cite{Chiou:2006qq}). Meanwhile, much effort has been made to study the Bianchi I LQC at the level of effective dynamics \cite{Chiou:2007dn,Chiou:2007sp,Chiou:2007mg}. Not only do the results of the effective dynamics agree with those in \cite{Chiou:2006qq} and \cite{Szulc:2008ar} but more intuitive pictures are obtained in the semiclassical approach. Thus, the study of effective dynamics can give us valuable insights into how and why the big bounces take place, even if the definite conclusion for the occurrence of big bounces has been drawn directly from the quantum theory.

However, the validity of the effective dynamics adopted in \cite{Chiou:2007dn,Chiou:2007sp,Chiou:2007mg} is only heuristic and remains to be justified. A systematic methodology to investigate the effective equations of motion for semiclassical states and the back-reaction resulting from quantum effects is on demand.

A mathematical framework has been developed in \cite{Bojowald:2005cw} to study the evolution of quantum systems in the realm close to classical behavior. In this framework, the behavior of a wave-function subject to a partial differential equation (Schr\"{o}dinger equation) is approximated by finitely many variables subject to a system of coupled but ordinary differential equations, giving a technically easier and more intuitive formulation to extract physical information of the semiclassical behavior amended by the back-reaction. This formalism has been used to address a variety of issues for quantum gravity and quantum cosmology \cite{Bojowald:2006ww,Bojowald:2006tm}. In particular, the bouncing scenario of LQC in the $k=0$ FRW model coupled to a free massless scalar has been analyzed from this perspective \cite{Bojowald:2006gr}. (An alternative approach for the effective equations of LQC in the $k=0$ FRW model was also studied in \cite{Taveras:2008ke}.)

In \cite{Bojowald:2006gr}, to apply the formulation of \cite{Bojowald:2005cw} to LQC in the $k=0$ FRW model, the constrained quantum system is viewed as an ordinary (unconstrained) quantum system by treating the scalar field $\phi$ as the time variable in the very beginning (referred to as ``Approach~I'' in this paper). Because of some technical difficulties, Approach I is inept for the Bianchi I model. In this paper, we propose a second approach (referred to as ``Approach II'') as a new framework to work with the effective equations of motion for the constrained quantum system that possesses an internal clock. In Approach II, the constrained quantum system is first treated as if there was no constraint and the formulation of \cite{Bojowald:2005cw} for unconstrained systems is adopted to obtain the effective equations of motion with respect to the coordinate time without directly treating $\phi$ as the time variable. On top of the finite system of effective equations of motion, the Hamiltonian constraint is then weakly imposed up to a certain order to further relate the finite set of variables. The technical and conceptual difficulties of the standard procedure to quantize a Hamiltonian constrained system are avoided in Approach II, the formulation of which can be intuitively understood as the natural extension of the classical Hamilton's equations subject to the classical Hamiltonian constraint. Additionally, this new approach has an extra merit that measurement of the internal time $\phi$ is on the equal footing as that of other observables (namely, the notions of uncertainty and so on for $\phi$ are retained), which is an attractive feature for the relativistic quantum mechanics that is still missing in the standard quantum theory for constrained systems. With the technical and conceptual virtues, Approach II per se could be used as a sound quantum theory for the constrained system if an internal time can be identified.

With the new formulation at hand, we investigate the effective equations of motion for the Wheeler-DeWitt (WDW) theory in the Bianchi I model with a free massless scalar. The results of Approach II agree with those obtained in the fully developed quantum theory in \cite{Chiou:2006qq} for the case without off-diagonal squeezing, thus evidencing the viability of the new formalism. Meanwhile, it is also revealed that back-reaction arises from the off-diagonal squeezing.

The next step is to apply Approach II to a simplified version of LQC in the Bianchi model. The resulting effective equations of motion affirm the bouncing scenario anticipated in \cite{Chiou:2006qq} and predicted in \cite{Chiou:2007dn,Chiou:2007mg}: The big bang is resolved and replaced by the big bounces, which take place up to three times, once in each diagonal direction, whenever the directional density $\varrho_I$ approaches the critical value $\varrho_{I,\mathrm{crit}}$ in the regime of Planckian density, but the precise value of $\varrho_{I,\mathrm{crit}}$ is modified by the quantum corrections. Moreover, the off-diagonal squeezing gives rise to further back-reaction, which couples the evolution in three diagonal directions.

Additionally, Approach II offers a language to describe the symmetry reduction in terms of effective equations of motion. In particular, it is notable that the anisotropic Bianchi I model admits the effective solution which exhibits isotropy, but such an effective solution is slightly different from that directly obtained in the isotropic model. This observation may help us understand more about the issue of symmetry reduction.

This paper is organized as follows. In \secref{sec:effective theory}, we first briefly review the mathematical framework developed in \cite{Bojowald:2005cw} and then introduce Approach I and formulate Approach II. In \secref{sec:effective WDW} and \secref{sec:effective LQC}, respectively, Approach II is adopted to investigate the WDW theory and simplified LQC in the Bianchi I model.\footnote{The reader may find Reference \cite{Chiou:2007dn} useful to serve as a quick review for the background knowledge needed.} The issue of symmetry reduction is studied in \secref{sec:symmetry reduction}. Finally, the results are summarized and discussed in \secref{sec:discussion}.

\section{Effective theory of quantum systems}\label{sec:effective theory}
In this section, we first briefly review the effective theory developed in \cite{Bojowald:2005cw} and then present two modified strategies --- Approach I and Approach II --- to study the constrained quantum system with an internal clock. Approach II is devised as a new formulation and its viability and virtues are discussed.

\subsection{Effective equations for ordinary quantum systems}\label{sec:ordinary QM system}
The effective theory of \cite{Bojowald:2005cw} is used to approximate the evolution of a wave-function subject to a partial differential equation (Schr\"{o}dinger equation) by a finite system of coupled but ordinary differential equations for finitely many variables.

For a given ordinary (unconstrained) quantum system\footnote{In this paper, we only consider quantum mechanical systems, not quantum field theories.} specified by a Hilbert space $\hilbert$ and a Hamiltonian $\hat{H}$, the evolution is given as a flow on $\hilbert$ by the Schr\"{o}dinger equation:
\be\label{eqn:Schrodinger eq}
-i\hbar\frac{d}{dt}\ket{\Psi(t)}=\hat{H}\ket{\Psi(t)}.
\ee

Equipped with the inner product $\inner{\cdot}{\cdot}$ in $\hilbert$, any operator $\hat{F}$ on $\hilbert$ defines a function $F:=\mean{\hat{F}}$ mapping $\ket{\Psi}\in\hilbert$ to the complex number $\inner{\Psi}{\hat{F}\vert\Psi}$. For two functions $F=\mean{\hat{F}}$ and $K=\mean{\hat{K}}$, the inner product of $\hilbert$ defines the Poisson bracket of $F$ and $K$:
\be\label{eqn:symplectic structure}
\{F,K\}=\frac{1}{i\hbar}\mean{[\hat{F},\hat{K}]},
\ee
which again maps $\ket{\Psi}\in\hilbert$ to a complex number. For example, $q:=\mean{\hat{q}}$ and $p:=\mean{\hat{p}}$ lead to $\{q,p\}=1$ from $[\hat{q},\hat{p}]=i\hbar$.

Let $(\hat{q}^i,\hat{p}_i)$ be the set of fundamental operators on $\hilbert$, we define the \emph{classical variables} as
\be
q^i:=\mean{\hat{q}^i},
\qquad
p_i:=\mean{\hat{p}_i},
\ee
and the \emph{quantum variables} of the $n$-th order as
\be
G^{i_1\dots i_n}:=
\mean{(\hat{x}^{i_1}-x^{i_1})\cdots(\hat{x}^{i_n}-x^{i_n})}_\mathrm{Weyl}
=\mean{(\hat{x}^{(i_1}-x^{(i_1})\cdots(\hat{x}^{i_n)}-x^{i_n)})}
\ee
associated with the fundamental operators $\{\hat{x}^i\}_{1\leq i\leq 2N}:=\{\hat{q}^k,\hat{p}_k\}_{1\leq k\leq N}$, where the subscript $\mathrm{Weyl}$ denotes symmetric ordering of the operators as the parenthesized indices represent the symmetrization: $A_{(i_1\dots i_n)}=1/n!\sum_{P}A_{i_{P(1)}\dots i_{P(n)}}$ for all permutations $P$ . These classical and quantum variables are bounded by Schwarz inequalities. A special case is the well-known uncertainly relation: $\mean{(\hat{q}-q)^2}\mean{(\hat{p}-p)^2}\geq
\mean{(\hat{q}-q)(\hat{p}-p)}_\mathrm{Weyl}^2+\hbar^2/4\geq\hbar^2/4$.

In terms of classical and quantum variables, the Schr\"{o}dinger equation \eqref{eqn:Schrodinger eq} govern by the Hamiltonian $\hat{H}$, which is usually taken to be the Weyl ordered operator $H(\hat{x}^i)_\mathrm{Weyl}$ with $H(x^i)$ being the classical counterpart, the quantum evolution can be equivalently described as the Hamiltonian flow:
\ba
\label{eqn:flow 1}
\dot{x}^i&=&\{x^i,H_Q\},\\
\label{eqn:flow 2}
\dot{G}^{i_1\dots i_n}&=&\{G^{i_1\dots i_n},H_Q\},
\ea
generated by the \emph{quantum Hamiltonian} defined as
\be\label{eqn:HQ}
H_Q:=\mean{H(\hat{x}^i)}_\mathrm{Weyl}=\mean{H(x^i+(\hat{x}^i)-x^i)}_\mathrm{Weyl}
=H(x^i)+
\sum_{n=2}^\infty
\frac{\partial^n H(x^i)}{\partial x^{(i_1}\cdots\partial x^{i_1)}}
\,G^{i_1\dots i_n}.
\ee

The reformulation gives the equations of motion in a classical form (i.e., a system of ordinary differential equations, albeit for infinitely many variables) and makes it possible to analyze the classical limit in a direct manner with quantum effects taken into account. While the classical variables represent the expectation values of the wave-function, the quantum variables carry the additional information (e.g., spreading, squeezing, etc.) of the wave-function around the peak. If the wave-function is highly coherent and thus sharply peaked around the semiclassical trajectory, as a good approximation, we can truncate the infinite set of coupled equations of motion to a finite set of differential equations by ignoring the higher order quantum variables. This offers a systematic method to study the effective equations of motion order by order. (For linear systems, such as the harmonic oscillator, the infinite set of coupled equations decouples into many sectors, each of which contains only finitely many variables.)

\subsection{Effective equations for constrained quantum systems: Approach I}\label{sec:Approach I}
Many quantum theories of our interest (such as general relativity and cosmology) are formulated as \emph{constrained} quantum systems, for which, however, the effective theory developed in \cite{Bojowald:2005cw} and outlined in \secref{sec:effective theory} cannot be directly applied. For a constrained system, the dynamical evolution is completely given by the \emph{Hamiltonian constraint} equation:
\be\label{eqn:Hamiltonian constraint}
\hat{H}\ket{\Psi}=0,
\ee
which dictates the correlations between physical variables, rather than the physical variables evolving with respect to a preferred time variable. Because the dynamics does not depend on the coordinate time explicitly, the effective theory for constrained quantum systems requires more considerations than that for ordinary quantum systems and its rigorous formulation is under development \cite{Bojowald:2008}.

Nevertheless, for the constrained system that possesses an \emph{internal clock}, there are ways to circumvent the difficulty arising from the absence of the explicit time variable, allowing us to analyze the dynamics in the same manner of an ordinary quantum system. To elaborate, let us consider a spatially homogeneous cosmological model with a scalar field $\phi$. By choosing the lapse function $N$ associated with the time coordinate $t'$ (i.e. $d\tau=Ndt'$ with $\tau$ being the proper time), the Hamiltonian constraint can be rescaled by $N$ to be of the form:
\be\label{eqn:H and Pphi}
\hat{H}=\hat{H}_0+\frac{\hat{p}_\phi^2}{2},
\ee
where $p_\phi$ is the conjugate momentum of $\phi$, satisfying $\{\phi,p_\phi\}=1$, and $\hat{H}_0$ is the rest part including the potential of $\phi$ (if any). The Hamiltonian constraint equation \eqnref{eqn:Hamiltonian constraint} then yields
\be\label{eqn:Approach I constraint}
\hat{p}_\phi\ket{\Psi}
\equiv -i\hbar\frac{\partial}{\partial\phi}\ket{\Psi}
=\pm (-2\hat{H}_0)^{1/2}\ket{\Psi}
=:\pm \hbar\,\hat{\Theta}^{1/2}\ket{\Psi},
\ee
where the sign $\pm$ is for \emph{positive/negative frequency} solutions (i.e. expanding/contracting solutions with respect to $\phi$), respectively. If the positive and negative frequency sectors do not interfere with each other (as in the case that $\phi$ is free of potential), we can restrict ourselves to one of them. Consequently, compared with the ordinary quantum system govern by \eqnref{eqn:Schrodinger eq}, the variable $\phi$ now plays the role of the time variable $t$ while $(\hbar^2\hat{\Theta})^{1/2}$ serves as the ordinary Hamiltonian.\footnote{\label{foot:positive definite}In general, $\hat{\Theta}$ may not be positive definite, but in the fully developed quantum theory, it turns out that the subspace of the negative eigenvalues of $\hat{\Theta}$ is nonphysical and thus the square root of $\hat{\Theta}$ is well-defined in the physical subspace on the positive spectrum of $\hat{\Theta}$. This can be easily understood if we rewrite \eqnref{eqn:Hamiltonian constraint} as $\hat{p}^2_\phi\ket{\Psi}=\hbar^2\hat{\Theta}\ket{\Psi}$, which excludes the negative spectrum of $\hat{\Theta}$ since $\hat{p}^2_\phi$ on the left-hand side is positive definite. More details can be found in \cite{Chiou:2006qq} for the Bicnchi I cosmology studied.} That is, the quantum evolution of a constrained system has been reformulated as an ordinary quantum system, with the scalar field treated as the ordinary time variable. This is the approach adopted in \cite{Bojowald:2006gr} to study the effective equations of motion for the quantum cosmology in the $k=0$ FRW model.

\subsection{Effective equations for constrained quantum systems: Approach II}\label{sec:Approach II}
For the quantum cosmology in the $k=0$ FRW model, the square root of $\hat{\Theta}$ remains polynomial of the fundamental operators and Approach I can be easily applied. Moreover, the Hamiltonian $\hat{\Theta}$ gives a linear system and as a consequence the equations of motions are exactly solvable \cite{Bojowald:2006gr}.

However, for more complicated cases, such as the quantum cosmology in the Bianchi I model, which will be the main focus of the rest of this paper, $\hat{\Theta}^{1/2}$ is no longer polynomial of the fundamental operators and because of the non-polynomiality the approximation scheme of Approach I is no longer under good control (as can be seen in \appref{app:WDW in Approach I}). To work around the technical difficulties, we innovate a second approach --- Approach II.

The idea of Approach II is to treat the constrained quantum system as if there was no constraint in the first place. The formulation outlined in \secref{sec:ordinary QM system} for ordinary quantum systems is then used to obtained the equations of motions with respect to the  coordinate time $t'$; that is, we define $H_Q$ via \eqnref{eqn:HQ} associated with $\hat{H}$ given by \eqnref{eqn:H and Pphi} and have
\ba
\label{eqn:Hamilton eq 1}
\frac{dx^i}{dt'}&=&\{x^i,H_Q\},\\
\label{eqn:Hamilton eq 2}
\frac{dG^{i_1\dots i_k}}{dt'}&=&\{G^{i_1\dots i_k},H_Q\}
\qquad\text{for }k=1,\,2,\cdots,n,
\ea
as the equations of motion for the classical and quantum variables up to the $n$-th order of our interest. On top of these finite differential equations, we then \emph{weakly} impose the Hamiltonian constraint as additional relations, by which the equations of motion abide, by demanding
\be\label{eqn:Approach II constraint}
\inner{\Psi}{(\hbar^2\hat{\Theta})^{k/2}|\Psi}=\inner{\Psi}{(\pm\hat{p}_\phi)^k|\Psi}
\qquad\text{for }k=1,\,2,\cdots,n,
\ee
up to the $n$-th order. If we only consider quantum variables up to the $n$-th order and neglect the higher order variables in \eqnref{eqn:Hamilton eq 2}, correspondingly, we should impose the constraint \eqnref{eqn:Approach II constraint} up to the same order. The $n$ constraints of \eqnref{eqn:Approach II constraint} will correlate the classical and quantum variables of the order $\leq n$ while higher order variables are consistently ignored. In the end, we eliminate $t'$ in favor of $\phi$ through the relation $d\phi/dt'$ given in \eqnref{eqn:Hamilton eq 1}. The resulting dynamics is in a form independent of the time coordinate $t'$ and predicts only the correlation between physical variables and $\phi$, which now serves as the \emph{internal time}.

Approach II, however, is not entirely equivalent to Approach I. For the WDW theory in the Bianchi I model, Approach II is studied in \secref{sec:effective WDW} and, for comparison, Approach I is included in \appref{app:WDW in Approach I}. The comparison shows that the two approaches yield the same result at the 1st order but already disagree at the 2nd order. When compared with the results of the fully developed quantum theory, it turns out that Approach II give sensible results while the order-by-order approximation scheme of Approach I is messed up due to the non-polynomiality.

Therefore, even though Approach II is only heuristically motivated, it gives a very good effective description of the quantum dynamics. If we take the viability of Approach II very seriously, we can even postulate Approach II as an alternative formulation for the fundamental theory of the constrained quantum system that possesses an internal clock. In order to be a viable formulation, in addition to giving the prediction very close to that of the standard treatment, Approach II has to be checked for self-consistency at least by two tests. First, for a given initial state which satisfies \eqnref{eqn:Approach II constraint}, after being evolved via \eqnref{eqn:Hamilton eq 1} and \eqnref{eqn:Hamilton eq 2}, the evolved state should satisfy \eqnref{eqn:Approach II constraint} again (up to the $n$-th order). Second, if we choose a different lapse function associated with a different time coordinate $t''$, $\hat{H}$ in \eqnref{eqn:H and Pphi} is rescaled accordingly and so is $H_Q$. The equations of motion given by \eqnref{eqn:Hamilton eq 1} and \eqnref{eqn:Hamilton eq 2} are then with the rescaled $H_Q$ and with respect to $t''$ but they should yield the same result (up to the order of our interest) when related to the internal time, regardless of different choice of the lapse function and coordinate time. If this is case, the coordinate time is really nothing but an auxiliary variable. Both tests are justified in \appref{app:self-consistency} but the justification is not completely stringent for generic cases.

In the rest of this paper, we focus on the quantum theory of cosmology in the Bianchi I model, which sets a benchmark to test the reliability of Approach II. As we will see, both for the WDW theory and simplified LQC, at least up to the 2nd order, the modified constraint \eqnref{eqn:Approach II constraint} simply correlates the constants of motion, and therefore the first test is trivially affirmed. Moreover, the studies of Approach II for the WDW theory and simplified LQC in the Bianchi I model give very sensible results compared to those obtained in the fully developed quantum theory. This is very instructive and suggests that Approach II, albeit not rigorously developed, could represent a sound quantum theory of its own even at the fundamental level.

In the standard procedure to quantize the constrained system, as the constraint is imposed via \eqnref{eqn:Hamiltonian constraint}, the physical states that satisfy the constraint are distributional if zero is part of the continuous spectrum of the constraint operator $\hat{H}$; in this case the physical states are not in the kinematic Hilbert space $\hilbert_\mathrm{kin}$ but instead in the dual space of $\hilbert_\mathrm{kin}$. Because of the distributional feature, the physical Hilbert space $\hilbert_\mathrm{phys}$ is rather difficult to construct and a variety of techniques have to be applied such as group averaging and refined algebraic quantization \cite{Marolf:2000iq,Marolf:1995cn} or the procedures performed in \cite{Ashtekar:2006uz,Ashtekar:2006wn} to define the physical inner product of $\hilbert_\mathrm{phys}$ by identifying a complete set of Dirac operators. On the other hand, when the scheme of Approach II is performed, we approximate the quantum system by \emph{finitely} many variables relevant to the accuracy of our measurement. Since we truncate the infinite degrees of freedom to finitely many variables and, accordingly, the Hamiltonian constraint is weakly imposed only to relate these finite degrees, the issue of the distributional property is completely gone and the kinematic Hilbert space remains the arena for the dynamics. The technical difficulties and conceptual obscurity of the standard treatment for constrained quantum systems are thus avoided in Approach II, whereby the evolution equations \eqnref{eqn:Hamilton eq 1} and \eqnref{eqn:Hamilton eq 2} subject to \eqnref{eqn:Approach II constraint} can be intuitively understood as the direct extension of the classical Hamilton's equations and the classical Hamiltonian constraint.

Additionally, Approach II has one extra merit as measurement of the internal time is regarded. Both in Approach I and in the standard treatment for the constrained quantum system, dynamical evolution is described as the correlation between physical variables and the internal time $\phi$, but, because of the aforementioned distributional property, $\phi$ plays a special role very different from other observables: In the resulting formalism, $\phi$ is treated as a pure parameter (c-number) as opposed to the other physical observables, which are treated as operators (q-numbers). In other words, the operator $\hat{\phi}$ is not defined in the physical Hilbert space ($\hat{p}_\phi$ is well-defined though) and therefore there are no notions of measurement, uncertainty and so on for $\phi$ unless inferred indirectly. In this regard, the philosophy of ``timeless'' formulation for relativistic quantum mechanics as advocated in \cite{Rovelli:1990jm,Rovelli:1991ni} is not fully realized in the standard treatment for constrained quantum systems. In Approach II, on the other hand, $\phi$ is put on the same footing as all other variables. Not only we can define the expectation value $\mean{\hat{\phi}}$, uncertainty $\Delta^2(\phi):= G^{\phi\phi}$ and so on for $\phi$, but the equations of motion \eqnref{eqn:Hamilton eq 1} and \eqnref{eqn:Hamilton eq 2} also tell us how they evolve with time.\footnote{More comments about the evolution of the variables involving $\phi$ can be found in the end of \secref{sec:WDW eom}.} The variable $\phi$ is no distinct from other observables except that it is also used as an internal time to label the evolution.\footnote{The reason why $\phi$, not other observables, serves as the internal time is because we take the square root in \eqnref{eqn:Approach II constraint} with respect to $\hat{p}_\phi^2$. It is explained in \appref{app:self-consistency} that the free massless scalar is special for being the internal time as the first test is concerned.} From the timeless perspective, Approaches II seems to give a more satisfactory framework such that all physical observables, including the internal time, correspond to measurements and every measurement yields uncertainty (see also \cite{Marolf:2002ve}). Compared with \eqnref{eqn:Hamiltonian constraint} for the standard treatment and \eqnref{eqn:Approach I constraint} for Approach I, the notable difference of Approach II is that the Hamiltonian constraint is only \emph{weakly} imposed as \eqnref{eqn:Approach II constraint} consistently ignores quantum fluctuations of higher orders. It seems to be the ignorance that retains the notions of measurement and uncertainty for $\phi$. This makes us suspect that the quantum Hamiltonian constraint \eqnref{eqn:Hamiltonian constraint} may turn out too restrictive and nonphysical, as it relates quantum fluctuations to every level of depth, which could be irrelevant after all as far as the physical measurements with intrinsic uncertainties are concerned.

Although some implications of Approach II remain speculative, the study of it can still teaches us valuable lessons for solving constrained quantum systems. In the following, we will apply Approach II to study the WDW theory and simplified LQC in the Bianchi I model as two sound examples and see explicitly how the ideas of Approach II materialize.

\section{Effective equations of motion for the WDW theory}\label{sec:effective WDW}
With the formulation of Approach II at hand, we are ready to study the effective equations of motion for the WDW theory in the Bianchi I model. The results, except those regarding measurement of $\phi$, will be shown to agree very well with those obtained from the fully developed quantum theory and thus the viability of Approach II is affirmed. On the other hand, Approach I is presented in \appref{app:WDW in Approach I} for the WDW theory and shown to be incompetent for the Bianchi I model.

\subsection{WDW theory in the Bianchi I model}\label{sec:WDW theory}
The phase space of the Bianchi I model is given by the diagonal triad
variables $p_I$ and connection variables $c^I$ for
$I=1,2,3$. The diagonal variables satisfy the canonical relation:
\be \{c^I,p_J\}=8\pi G\gamma\,\delta^I_J.
\ee
In the presence of a free massless scalar field $\phi({\vec x},t)=\phi(t)$,
(which is independent of the spatial coordinates with homogeneity assumed),
the classical dynamics is govern
by the Hamiltonian constraint:
\ba\label{eqn:original H}
H&=&H_\mathrm{grav}+H_{\phi}\nn\\
&=&-\frac{1}{8\pi G\gamma^2\sqrt{{p_1p_2p_3}}}
\left({c^2c^3}{p_2p_3}+{c^1c^3}{p_1p_3}+{c^1c^2}{p_1p_2}\right)
+\frac{p_{\phi}^2}{2\sqrt{{p_1p_2p_3}}},
\ea
where $p_\phi$ is the conjugate momentum of $\phi$ and has the canonical relation:
\be
\{\phi,p_\phi\}=1.
\ee

We can rescale the Hamiltonian by choosing the lapse function $N=\sqrt{p_1p_2p_3}$
associated with the new time variable $t'$ via $dt'=(p_1p_2p_3)^{-1/2}d\tau$. The new (rescaled) Hamiltonian constraint reads as
\be\label{eqn:H}
H=-\frac{1}{8\pi G\gamma^2}
\left({c^2c^3}{p_2p_3}+{c^1c^3}{p_1p_3}+{c^1c^2}{p_1p_2}\right)
+\frac{p_\phi^2}{2}.
\ee

In the Wheeler-DeWitt (WDW) theory, the standard Schr\"{o}dinger quantization
is adopted. The kinematic Hilbert space is
$\hilbert^\mathrm{WDW}_\mathrm{kin}\!=L^2(\mathbb{R}^3,d^3p)\otimes L^2(\mathbb{R},d\phi)$.
The variables $c^I$, $p_I$, $\phi$, $p_\phi$ are lifted to the operators $\hat{c}^I$, $\hat{p}_I$,
$\hat{\phi}$ and $\hat{p}_\phi$. In particular, $\hat{c}_I$ and $\hat{p}_\phi$ are
represented in $\hilbert^\mathrm{WDW}_\mathrm{kin}$ as differential operators:
$\hat{c}^I\rightarrow i\hbar\,8\pi G\gamma\frac{\partial}{\partial p_I}$,
$\hat{p}_\phi\rightarrow -i\hbar\frac{\partial}{\partial\phi}$.
We also define the hermitian operators:
\be
\hat{K}^I=\frac{1}{2}\left(\hat{c}^I\hat{p}_I+\hat{p}_I\hat{c}^I\right)
\ee
and correspondingly define the Hamiltonian
operator as
\be\label{eqn:H hat}
\hat{H}=-\frac{1}{8\pi G\gamma^2}
\left(\hat{K}^2\hat{K}^3+\hat{K}^1\hat{K}^3+\hat{K}^1\hat{K}^2\right)
+\frac{\hat{p}_\phi^2}{2}
=:-\frac{1}{2}\left\{\hbar^2\hat{\Theta}-\hat{p}_\phi^2\right\},
\ee
by fixing the ordering such that $\hat{H}$ is hermitian.\footnote{The WDW theory constructed here is very similar to that in \cite{Chiou:2006qq}. In particular, $\hat{\Theta}\approx\underline{\Theta}$ ($\underline{\Theta}$ defined in \cite{Chiou:2006qq}) with very tiny difference due to the different choices of ordering. $\hat{\Theta}$ is not positive definite but the square root of $\hat{\Theta}$ is well-defined in the physical subspace. Also see \footref{foot:positive definite}.}

\subsection{Classical and quantum variables}
\label{sec:WDW variables}
In order to apply Approach II to derive the effective equations of motion, we define classical variables as\footnote{For technical convenience, instead of the canonical variables $(p_I,c^J)$, we choose $(p_I,K^J)$ as the fundamental variables. For reference, the Poisson brackets based on the canonical variables are listed in \appref{app:standard variables}.}
\ba
p_I&:=&\mean{\hat{p}_I},\qquad
K^I:=\mean{\hat{K}^I},\\
\label{eqn:phi pphi}
p_\phi&:=&\mean{\hat{p}_\phi},\qquad
\phi:=\mean{\hat{\phi}},
\ea
which, by \eqnref{eqn:symplectic structure}, satisfy the Poisson brackets:
\ba
\{p_I,K^J\}=-8\pi G\gamma\delta_I^J p_I,\qquad
\{\phi,p_\phi\}=1,\qquad
\{p_I,p_J\}=\{K^I,K^J\}=0.
\ea

The associated quantum variables of the 2nd order are also defined:
\ba
G^{(n=2)}:\qquad\quad
G_{IJ}&:=&\mean{(\hat{p}_I-\mean{\hat{p}_I})(\hat{p}_J-\mean{\hat{p}_J})},\nn\\
G^{IJ}&:=&\mean{(\hat{K}^I-\mean{\hat{K}^I})(\hat{K}^J-\mean{\hat{K}^J})},\nn\\
G^I_J&:=&\mean{(\hat{K}^I-\mean{\hat{K}^I})(\hat{p}_J-\mean{\hat{p}_J})}_\mathrm{Weyl},\\
G^{\phi,(n=2)}:\qquad
G_{p_\phi p_\phi}&:=&\mean{(\hat{p}_\phi-\mean{\hat{p}_\phi})(\hat{p}_\phi-\mean{\hat{p}_\phi})},\nn\\
G^{\phi\phi}&:=&\mean{(\hat{\phi}-\mean{\hat{\phi}})(\hat{\phi}-\mean{\hat{\phi}})},\nn\\
G^\phi_{p_\phi}&:=&\mean{(\hat{\phi}-\mean{\hat{\phi}})(\hat{p}_\phi-\mean{\hat{p}_\phi})}_\mathrm{Weyl}.
\ea
We have $\{\phi,G^{\phi,(n=2)}\}=\{p_\phi,G^{\phi,(n=2)}\}=0$, but on the other hand,
as opposed to the case for canonical variables $(\hat{p}_I,\hat{c}^J)$,
the commutator $[\hat{p}_I,\hat{K}^J]$ is not a constant and consequently the quantum variables do not commute with the classical variables but, by \eqnref{eqn:symplectic structure}, yield the relations:
\ba
&&\{p_I,G_{JK}\}=0,\qquad
\{p_I,G^{JK}\}=-8\pi G\gamma\left(\delta_I^J G_J^K+\delta_I^K G_K^J\right),\nn\\
&&\{p_I,G^J_K\}=-8\pi G\gamma \delta_I^J G_{JK},
\ea
and
\ba
&&\{K^I,G_{JK}\}=-8\pi G\gamma\left(\delta^I_J +\delta^I_K\right)G_{JK},\qquad
\{K^I,G^{JK}\}=0,\nn\\
&&\{K^I,G^J_K\}=-8\pi G\gamma \delta^I_K G^J_K.
\ea

We also have
\be
\{G^{\phi\phi},G^\phi_{p_\phi}\}=2G^{\phi\phi},\quad
\{G^{\phi\phi},G_{p_\phi p_\phi}\}=4G^\phi_{p_\phi},\quad
\{G^\phi_{p_\phi},G_{p_\phi p_\phi}\}=2G_{p_\phi p_\phi},
\ee
and
\ba\label{eqn:commutators of G and G}
&&\{G^I_J,G^K_L\}=(8\pi G\gamma)\left(\delta^I_L G^K_{JI}-\delta^K_J G^I_{LK}\right),
\quad
\{G^{IJ},G_{KL}\}=4 (8\pi G\gamma) \delta^{(I}_{(K} G^{J)}_{L)I},\nn\\
&&\{G^{IJ},G^K_L\}=2 (8\pi G\gamma) \delta^{(I}_{L} G^{J)K}_{I},\qquad
\{G_{IJ},G^K_L\}=-2(8\pi G\gamma) \delta^{K}_{(I} G_{J)LK},\nn\\
&&\{G_{IJ},G_{KL}\}=\{G^{IJ},G^{KL}\}=0,
\ea
where the 3rd order quantum variables ($G^{(n=3)}$) are defined in the obvious
manner (e.g.,
$G^I_{JK}:=\mean{(\hat{K}^I-\mean{\hat{K}^I})
(\hat{p}_J-\mean{\hat{p}_J})(\hat{p}_K-\mean{\hat{p}_K})}_\mathrm{Weyl}$).
The Schwartz inequality leads to the relations:
\be\label{eqn:Schwartz 1}
G^{\phi \phi}G_{p_\phi p_\phi}\geq {G^\phi_{p_\phi}}^2+\frac{\hbar^2}{4}
\ee
and
\ba
\label{eqn:Schwartz 2}
G^{II}G^{JJ}&\geq& {G^{IJ}}^2,\\
\label{eqn:Schwartz 3}
G_{II}G_{JJ}&\geq& {G_{IJ}}^2,\\
\label{eqn:Schwartz 4}
G^{II}G_{JJ}&\geq& {G^I_J}^2+\frac{1}{4}\abs{\mean{[\hat{K}^I,\hat{p}_J]}}^2
={G^I_J}^2+\frac{(8\pi G\hbar\gamma)^2}{4}\delta^I_Jp_I^2.
\ea

\subsection{Effective equations of motion}\label{sec:WDW eom}
Corresponding to \eqnref{eqn:H hat}, the quantum Hamiltonian defined in \eqnref{eqn:HQ} reads as
\be
H_Q=-\frac{1}{8\pi G\gamma^2}\left(K^2K^3+K^1K^3+K^1K^2+G^{23}+G^{13}+G^{12}\right)
+\frac{p_\phi^2}{2}+\frac{G_{p_\phi p_\phi}}{2}.
\ee
According to \eqnref{eqn:Hamilton eq 1} and the Poisson brackets listed in \secref{sec:WDW variables}, the equations of motion for the classical variables are given by
\ba
\label{eqn:wdw eom1}
\frac{dK^I}{dt'}&=&\{K^I,H_Q\}=0\quad\Rightarrow\quad K^I:=8\pi G\gamma\hbar\,{\calK}^I \text{ are constant},\\
\label{eqn:wdw eom2}
\frac{dp_1}{dt'}&=&\{p_1,H_Q\}=\gamma^{-1}\left\{(K^2+K^3)p_1+G_1^2+G_1^3\right\},\\
\label{eqn:wdw eom3}
\frac{d\phi}{dt'}&=&\{\phi,H_Q\}=p_\phi,\\
\label{eqn:wdw eom4}
\frac{dp_\phi}{dt'}&=&\{\phi,H_Q\}=0\quad\Rightarrow\quad
p_\phi:=\hbar\sqrt{8\pi G}{\calK}_\phi \text{ is constant},
\ea
where we define the dimensionless constants ${\calK}^I$ and ${\calK}_\phi$ for convenience. We also have the equations for $dp_2/dt'$ and $dp_3/dt'$ in the cyclic rearrangement of \eqnref{eqn:wdw eom2}.\footnote{The obvious cyclic repetition will not be mentioned again hereafter.} Similarly, \eqnref{eqn:Hamilton eq 2} gives the equations of motion for the quantum variables:
\ba
\label{eqn:wdw eom5}
\frac{dG^{IJ}}{dt'}&=&\{G^{IJ},H_Q\}=0\quad\Rightarrow\quad G^{IJ} \text{ are constant},\\
\label{eqn:wdw eom6}
\frac{dG_{12}}{dt'}&=&\{G_{12},H_Q\}=\gamma^{-1}G_{12}\left(K^1+K^2+2K^3\right)+\gamma^{-1}{\cal O}(G^{(n=3)}),\\
\label{eqn:wdw eom7}
\frac{dG_{11}}{dt'}&=&\{G_{11},H_Q\}=\gamma^{-1}G_{11}\left(2K^2+2K^3\right)+\gamma^{-1}{\cal O}(G^{(n=3)}),\\
\label{eqn:wdw eom8}
\frac{dG^I_1}{dt'}&=&\{G^I_1,H_Q\}=\gamma^{-1}G^I_1\left(K^2+K^3\right)+\gamma^{-1}{\cal O}(G^{(n=3)}),
\ea
and
\ba
\label{eqn:wdw eom9}
\frac{dG_{p_\phi p_\phi}}{dt'}&=&\{G_{p_\phi p_\phi},H_Q\}=0\quad\Rightarrow\quad G_{p_\phi p_\phi} \text{ is constant},\\
\label{eqn:wdw eom10}
\frac{dG^\phi_{p_\phi}}{dt'}&=&\{G^\phi_{p_\phi},H_Q\}=G_{p_\phi p_\phi},\\
\label{eqn:wdw eom11}
\frac{dG^{\phi\phi}}{dt'}&=&\{G^{\phi\phi},H_Q\}=2G^\phi_{p_\phi}.
\ea

Note that \eqnref{eqn:wdw eom3} and \eqnref{eqn:wdw eom4} imply that $\phi$ is a
monotonic function of $t'$ and thus can be used as an internal time.
If the wave-function is sharply peaked, ${\cal O}(G^{(n=3)})$ terms are negligible,
and, in terms of the internal time $\phi$, the general solutions to the above equations of motion are given by
\ba
\label{eqn:sol for GIJ}
G^{IJ}(\phi)&=&g^{IJ},\\
G_{12}(\phi)&\approx&g_{12}\,e^{\sqrt{8\pi G}\left(\frac{{\calK}^1+{\calK}^2+2{\calK}^3}{{\calK}_\phi}\right)(\phi-\phi_0)},\\
\label{eqn:sol for GII}
G_{11}(\phi)&\approx&g_{11}\,e^{\sqrt{8\pi G}\left(\frac{2{\calK}^2+2{\calK}^3}{{\calK}_\phi}\right)(\phi-\phi_0)},\\
G_1^I(\phi)&\approx&g_1^I\,e^{\sqrt{8\pi G}\left(\frac{{\calK}^2+{\calK}^3}{{\calK}_\phi}\right)(\phi-\phi_0)},\\
\label{eqn:sol for pI}
p_1(\phi)&\approx&\left(p_{1,0}+\frac{g_1^2+g_1^3}{\gamma\hbar\sqrt{8\pi G}{\calK}_\phi}(\phi-\phi_0)\right)
e^{\sqrt{8\pi G}\left(\frac{{\calK}^2+{\calK}^3}{{\calK}_\phi}\right)(\phi-\phi_0)},
\ea
and
\ba
\label{eqn:sol for Gpp}
G_{p_\phi p_\phi}(\phi)&=&g_{p_\phi p_\phi},\\
\label{eqn:sol for Gpphi}
G_{p_\phi}^\phi(\phi)&=&g^\phi_{p_\phi}+\frac{g_{p_\phi p_\phi}}{\hbar\sqrt{8\pi G}{\calK}_\phi}(\phi-\phi_0),\\
\label{eqn:sol for Gphiphi}
G^{\phi\phi}(\phi)&=&g^{\phi\phi}+\frac{2g^\phi_{p_\phi}}{\hbar\sqrt{8\pi G}{\calK}_\phi}(\phi-\phi_0)
+\frac{g_{p_\phi p_\phi}}{(\hbar\sqrt{8\pi G}{\calK}_\phi)^2}(\phi-\phi_0)^2,
\ea
where $p_{I,\,0}$, $g^{IJ}$, $g_{IJ}$, $g^{\phi\phi}$, etc. are all constants,
which satisfy the inequalities
\ba
\label{eqn:ineq 1}
&&g^{II}g^{JJ}\geq{g^{IJ}}^2,\quad
g_{II}g_{JJ}\geq{g_{IJ}}^2,\quad \text{for }\forall\ I,J,\\
\label{eqn:ineq 2}
&&g^{II}g_{JJ}\geq{g^I_J}^2,\quad \text{for } I\neq J,\\
\label{eqn:ineq 3}
&&g^{11}g_{11}\geq{g^1_1}^2+\frac{(8\pi G\hbar\gamma)^2}{4}
\left(p_{1,\,0}+\frac{g_1^2+g_1^3}{\gamma\hbar\sqrt{8\pi G}{\calK}_\phi}(\phi-\phi_0)\right)^2,
\ea
and
\be\label{eqn:ineq 4}
g^{\phi\phi}g_{p_\phi p_\phi}\geq {g^\phi_{p_\phi}}^2 + \frac{\hbar^2}{4}
\ee
according to \eqnref{eqn:Schwartz 1}--\eqnref{eqn:Schwartz 4}.

Later in \secref{sec:Hamiltonian constraint}, we will show that the Hamiltonian constraint
gives the relation \eqnref{eqn:K} for the dimensionless constants ${\calK}^I$ and ${\calK}^\phi$, which agrees with the classical counterpart with small corrections.
Consequently, compared with the classical solution: $p_1(\phi)=p_{1,0}\exp(\sqrt{8\pi G}(\phi-\phi_0)(\calK^2+\calK^3)/\calK_\phi)$, \eqnref{eqn:sol for pI} shows that
the expectation value of each $\hat{p}_I$ closely follows the classical trajectory but will receive back-reaction if the off-diagonal squeezing is nonzero
(i.e. $g_1^2,g_1^3\neq0$ etc).\footnote{In this paper, \emph{squeezing} refers to the correlation between $p_I$ and $K^J$ (i.e. $G_I^J$) or between $\phi$ and $p_\phi$ (i.e. $G^\phi_{p_\phi}$). The phrase ``off-diagonal'' refers to $G_{IJ}$, $G^{IJ}$ and $G^I_J$ with $I\neq J$ while ``diagonal'' to those with $I=J$.} Furthermore, in the case without off-diagonal squeezing (i.e., $g_1^2,g_1^3=0$ etc), according to \eqnref{eqn:sol for GII} and \eqnref{eqn:sol for pI}, the relative spread for each $p_I$ remains constant: i.e., $\Delta p_I/p_I:=\sqrt{G_{II}}/p_I=\sqrt{g_{II}}/p_{I,\,0}$.

The fully developed quantum theory of the Bianchi I WDW cosmology has been studied in Section VI and Appendix A of \cite{Chiou:2006qq}. The semiclassical state constructed in (A.1) of \cite{Chiou:2006qq} is with constant spreads in $K_I$ (i.e. $G^{II}=(8\pi \gamma\hbar)^2\sigma_I^2$), zero off-diagonal correlations (i.e. $G_{IJ}=0$ for $I\neq J$) and zero squeezing (i.e. $G^J_I=0$). The evolution of this state is given by Equations (6.19), (6.21) and (6.22) in \cite{Chiou:2006qq}, which indicate that the peak of the wave-function follows the classical trajectory with constant relative spreads. That said, the behaviors of \eqnref{eqn:sol for GIJ}--\eqnref{eqn:sol for pI} conform to the fully developed quantum theory at least for the case studied in \cite{Chiou:2006qq}. The results here also instructs us to construct a new type of coherent states with nonzero squeezing and off-diagonal correlations by generalizing (A.1) in \cite{Chiou:2006qq} to see if the evolution in the fully developed quantum theory receives the back-reaction as suggested in \eqnref{eqn:sol for pI}. We expect the answer to be affirmative. By contrast, as shown in \appref{app:WDW in Approach I}, the scheme of Approach I does not give the same results as those in \cite{Chiou:2006qq} and is thus inept for the Bianchi I model.

Meanwhile, as mentioned in \secref{sec:Approach II}, the notions for the measurements involving $\phi$ are well-posed in Approach II and furthermore how they evolve can be answered. According to \eqnref{eqn:sol for Gpp}--\eqnref{eqn:sol for Gphiphi}, the spread of $p_\phi$ remains constant, but the squeezing and the spread of $\phi$ grow with $\phi$. This is a well-known fact as for the quantum mechanics of a free particle, which has a strictly growing spread. For unbounded systems, we cannot expect to have a valid semiclassical approximation for all times, but the semiclassical treatment is still reasonable for limited amounts of time (see Example 3 in \cite{Bojowald:2005cw} for more details). This implies that treating $\phi$ as a classical clock is only valid for a limited period of time. This problem can be mended by adding a mass term to $\phi$, which then behaves as a harmonic oscillator instead of a free particle. In the presence of the mass term, as for a harmonic oscillator, the wave-function always admits bounded spread and thus the semiclassicality can be sustained forever. Even though $\phi$ is no longer a monotonic function of $t'$, it can still be used \emph{locally} as a classical clock for all times. It seems to suggests that only the field with oscillatory behavior of harmonic type can be viewed as a classical clock. This observation might shed light on the problem of time (see \cite{Isham:1992ms} for an overview) by offering insight into why in reality the passage of time is always measured in terms of oscillatory signals.

\subsection{Hamiltonian constraint}\label{sec:Hamiltonian constraint}
In addition to solving the effective equations of motion, we still have to impose the Hamiltonian constraint \eqnref{eqn:Approach II constraint}, which will give further relations to relate the constants of motion used to parameterize the semiclassical solutions. Since we have solved the equations of motion up to the order of $G^{(n=2)}$ and $G^{\phi,(n=2)}$, accordingly, we only consider $k=1,2$ in \eqnref{eqn:Approach II constraint}. (The Hamiltonian constraint for $k\geq3$ will further constrain the higher order terms $G^{(\phi,n\geq3)}$ with $G^{(n\geq3)}$ and other parameter constants.)

For $k=2$, \eqnref{eqn:Approach II constraint} yields
\ba\label{eqn:constraint 1}
\mean{\hat{p}_\phi^2}&=&\frac{1}{4\pi G\gamma^2}
\left\{\mean{\hat{K}^2\hat{K}^3}+\mean{\hat{K}^1\hat{K}^3}
+\mean{\hat{K}^1\hat{K}^2}\right\}\nn\\
&=&\frac{1}{4\pi G\gamma^2}\left\{K^2K^3+K^1K^3+K^1K^2+G^{23}+G^{13}+G^{12}\right\}\nn\\
&=&16\pi G\hbar
\left\{{\calK}^2{\calK}^3+{\calK}^1{\calK}^3+{\calK}^1{\calK}^2\right\}
+\frac{1}{4\pi G\gamma^2}\left\{g^{23}+g^{13}+g^{12}\right\}.
\ea
For $k=1$, by defining ${\theta}(K^I):=(K^2K^3+K^1K^3+K^1K^2)$ and $\sqrt{\theta}(K^I):=(K^2K^3+K^1K^3+K^1K^2)^{1/2}$,
\eqnref{eqn:Approach II constraint} yields
\ba\label{eqn:constraint 2}
&&\mean{\hat{p}_\phi}=\frac{1}{\sqrt{4\pi G\gamma^2}}
\mean{\left(\hat{K}^2\hat{K}^3+\hat{K}^1\hat{K}^3+\hat{K}^1\hat{K}^2\right)^{1/2}}\nn\\
&=&\frac{1}{\sqrt{4\pi G\gamma^2}}\mean{\sqrt{\theta}(K^I+(\hat{K}^I-K^I))}\nn\\
&=&\frac{1}{\sqrt{4\pi G\gamma^2}}
\left\{\sqrt{\theta}(K^I)+\sum_{n=2}^\infty\sum_{I_1,\dots, I_n}\frac{1}{n!}\,
\frac{\partial^n\sqrt{\theta}}{\partial K^{I_1}\dots\partial K^{I_n}}\,G^{I_1I_2\dots I_n}\right\}\nn\\
&=&\frac{1}{\sqrt{4\pi G\gamma^2}}\Biggl\{\sqrt{\theta}(K^I)
-\frac{(K^2+K^3)^{2}}{8\,\theta(K^I)^{3/2}}\,G^{11}
-\frac{(K^1+K^3)^{2}}{8\,\theta(K^I)^{3/2}}\,G^{22}
-\frac{(K^1+K^2)^{2}}{8\,\theta(K^I)^{3/2}}\,G^{33}
\\
&&
+\frac{\theta(K^I)-{K^1}^2}{4\,\theta(K^I)^{3/2}}\,G^{23}
+\frac{\theta(K^I)-{K^2}^2}{4\,\theta(K^I)^{3/2}}\,G^{13}
+\frac{\theta(K^I)-{K^3}^2}{4\,\theta(K^I)^{3/2}}\,G^{12}
+\sqrt{\theta}^{1-n}{\cal O}(G^{(n\geq3)})\Biggr\},\qquad\nn
\ea
and consequently
\be
\mean{\hat{p}_\phi}^2
\approx\frac{1}{4\pi G\gamma^2}
\left\{\theta(K^I)-\frac{(K^2+K^3)^2}{4\,\theta(K^I)}\,G^{11}-\cdots
+\frac{\theta(K^I)-{K^1}^2}{2\,\theta(K^I)}\,G^{23}
+\cdots\right\},
\ee
which can be rewritten as
\ba\label{eqn:K}
\calK_\phi^2&\approx&2
\left(\calK^2\calK^3+\calK^1\calK^3+\calK^1\calK^2\right)
+\frac{2}{(8\pi G\hbar\gamma)^2}
\Biggl\{-\frac{1}{4\,}\frac{({\calK}^2+{\calK}^3)^2}
{({\calK}^2{\calK}^3+{\calK}^1{\calK}^3+{\calK}^1{\calK}^2)}\,g^{11}-\cdots\nn\\
&&\qquad\qquad\qquad+
\frac{({\calK}^2{\calK}^3+{\calK}^1{\calK}^3+{\calK}^1{\calK}^2)-{{\calK}^1}^2}
{2({\calK}^2{\calK}^3+{\calK}^1{\calK}^3+{\calK}^1{\calK}^2)}\,g^{23}
+\cdots\Biggr\},
\ea
closely in agreement with the classical counterpart $\calK_\phi^2=2\left(\calK^2\calK^3+\calK^1\calK^3+\calK^1\calK^2\right)$ with small corrections arising from the quantum fluctuations.

Furthermore, \eqnref{eqn:constraint 1} and \eqnref{eqn:constraint 2} together give the spread of $p_\phi$:
\ba\label{eqn:K 2}
&&(\Delta p_\phi)^2:=\mean{\hat{p}_\phi^2}-\mean{\hat{p}_\phi}^2\equiv G_{p_\phi p_\phi}=g_{p_\phi p_\phi}\nn\\
&\approx&\frac{(4\pi G\gamma^2)^{-1}}{({\calK}^2{\calK}^3+{\calK}^1{\calK}^3+{\calK}^1{\calK}^2)}
\biggl\{
\frac{1}{4}\left[({\calK}^2+{\calK}^3)^2g^{11}+({\calK}^1+{\calK}^3)^2g^{22}+
({\calK}^1+{\calK}^2)^2g^{33}\right]\nn\\
&&\qquad+\frac{1}{2}\left[{{\calK}^1}^2g^{23}
+{{\calK}^2}^2g^{13}+{{\calK}^3}^2g^{12}\right]
\biggr\}
+\frac{(4\pi G\gamma^2)^{-1}}{2}\left(g^{23}+g^{13}+g^{12}\right).
\ea
In the case without off-diagonal spreads in $K^I$ (i.e. $g^{IJ}=0$ for $I\neq J$), this leads to
\be
\frac{(\Delta p_\phi)^2}{p_\phi^2}\approx
\frac{1}{{\calK}_\phi^4}\left\{({\calK}^2+{\calK}^3)^2\sigma_1^2+({\calK}^1
+{\calK}^3)^2\sigma_2^2+({\calK}^1+{\calK}^2)^2\sigma_3^2\right\},
\ee
where the dimensionless constants $\sigma_I^2:=G^{II}/(8\pi G\gamma\hbar)^2$ are the spreads for $K^I/(8\pi G\gamma\hbar)$.
This result concurs with Equation (6.24) of \cite{Chiou:2006qq} for the WDW semiclassical state except the mismatch of an overall factor 2 due to the approximation method used in \cite{Chiou:2006qq} to evaluate $\Delta p_\phi$. The agreement further strengthens the viability of Approach II.

\section{Effective equations of motion for simplified LQC}\label{sec:effective LQC}
The viability of Approach II has been tested for the WDW theory. In this section, we will investigate the effective equations of motion for the simplified LQC in the Bianchi I model along the line of Approach II. The results are expected to be in accordance with those obtained in the semiclassical approach \cite{Chiou:2007dn,Chiou:2007mg} amended with back-reaction resulting from quantum fluctuations.

\subsection{Simplified LQC in the Bianchi I model}\label{sec:LQC simplified LQC}
The detailed construction for the Hamiltonian operator corresponding to \eqnref{eqn:original H} in the fully developed LQC can be found in \cite{Chiou:2006qq}. The LQC quantization is implemented with two main sources of quantum corrections: First, the connection variables $c^I$ do not exist and should be replace by holonomies; second, the inverse triads $p_I^{-1/2}$, upon quantization, receive quantum corrections via the so-called Thiemann's trick. It is realized that the corrections on the inverse triads are less important when the wave-function evolves in the semiclassical realm. To simplify the quantum theory, the Thiemann's trick is simply ignored; i.e., the Hamiltonian constraint \eqnref{eqn:original H} is trivially rescaled to \eqnref{eqn:H} before quantization.\footnote{This the \emph{simplified} LQC studied in \cite{Ashtekar:2007em,Corichi:2007am} for the isotropic case, where the simplification leads to the exact solvability.}

Therefore, we start with the rescaled Hamiltonian \eqnref{eqn:H} and take the prescription to replace $c^I$ by $\sin(\mubar_Ic^I)/\mubar_I$ by introducing discreteness variables $\mubar_I$. The rescaled Hamiltonian is modified to
\be\label{eqn:H bar}
\bar{H}=
-\frac{1}{8\pi G \gamma^2}
\left\{
\frac{\sin(\mubar_2c^2)\sin(\mubar_3c^3)}{\mubar_2\mubar_3}p_2p_3+
\text{cyclic terms}
\right\}
+\frac{p_\phi^2}{2}.
\ee
There is a variety of possibilities to implement the $\mubar$ discreteness and two well-motivated constructions (referred to as the ``$\mubar$ scheme'' and ``$\mubar'$ scheme'') are discussed in great detail in \cite{Chiou:2007mg}. The $\mubar'$ scheme has a better scaling property that the quantum dynamics based on it is invariant under different choice of the fiducial cell. Unfortunately, the quantum theory of the $\mubar'$ scheme is much difficult and has yet to be constructed. In this paper, we study the $\mubar$-scheme quantization described in \cite{Chiou:2006qq} with the aforementioned simplification. In the $\mubar$-scheme, $\mubar_I$ are prescribed as
\be
\mubar_1=\sqrt{\frac{\Delta}{p_1}}\;,
\qquad
\mubar_2=\sqrt{\frac{\Delta}{p_2}}\;,
\qquad
\mubar_3=\sqrt{\frac{\Delta}{p_3}}\;,
\ee
where $\Delta$ is the \emph{area gap} in the full theory of LQG and $\Delta=\frac{\sqrt{3}}{2}(4\pi\gamma\Pl^2)$ for the standard choice (but other choices are also possible) with $\Pl=\sqrt{G\hbar}$ being the Planck length.

The kinematic Hilbert space for LQC is given by $\hilbert_\mathrm{kin}^\mathrm{LQC}=L^2(\mathbb{R}^3_\mathrm{Bohr},d^3p_\mathrm{Bohr})
\otimes L^2(\mathbb{R},d\phi)$, where $\phi$ is in the ordinary Schr\"{o}dinger representation while $\vec{p}$ is in the ``polymer representation'', which reflects the very feature of LQC that the connection operators $\hat{c}^I$ cannot be defined. Our task now is to promote \eqnref{eqn:H bar} to a well-defined quantum operator in $\hilbert_\mathrm{kin}^\mathrm{LQC}$. Following the strategy used in \cite{Chiou:2006qq}, we define the dimensionless \emph{affine variables} $v_I$ to satisfy
\be
(4\pi\gamma\Pl^2)\,\mubar_I\frac{\partial}{\partial p_I}=\frac{\partial}{\partial v_I},
\ee
which yields
\be\label{eqn:p and v}
\frac{p_I}{\mubar_I}=\frac{p_I^{3/2}}{\Delta^{1/2}}=6\pi\gamma\Pl^2\, v_I.
\ee
In $L^2(\mathbb{R}^3_\mathrm{Bohr},d^3p_\mathrm{Bohr})
=L^2(\mathbb{R}^3_\mathrm{Bohr},d^3v_\mathrm{Bohr})$, we can then quantize $\exp(\pm i\mubar_I c^I/2)$ as the translation operators
\be\label{eqn:exp hat}
\widehat{e^{\pm\frac{i}{2}\mubar_1 c^1}}\psi(\vec{v}):=
e^{\mp\frac{\partial}{\partial v_1}}\psi(\vec{v})=\psi(v_1\mp1,v_2,v_3)
\ee
and so on in the $\vec{v}$-representation.\footnote{If the Hilbert space was in the ordinary Schr\"{o}dinger representation, $\hat{c}^I$ would be well-defined and represented as  $\hat{c}^I\rightarrow i\hbar\,8\pi G\gamma \frac{\partial}{\partial p_I}$ in the $\vec{p}$-representation. In the ordinary quantum mechanics, $\hat{p}\rightarrow-\frac{i}{\hbar}\frac{d}{dx}$ and $e^{i a \hat{p}/\hbar}\psi(x)=\psi(x+a)$, but we cannot simply take $e^{i f(x)\hat{p}/\hbar}\rightarrow e^{f(x)\frac{d}{dx}}$ because it is not unitary in general. Here, by contrast, since $\vec{p}$ is in the polymer representation, the same problem does not occur. The operators defined in \eqnref{eqn:exp hat} are unitary in $L^2(\mathbb{R}^3_\mathrm{Bohr},d^3v_\mathrm{Bohr})$. (Also see Footnote 10 in \cite{Chiou:2006qq}).}
Here, the wave function $\psi(\vec{v})$ is related to $\psi(\vec{p})$ via
\be
\abs{\psi(\vec{v})}^2 d^3v=\abs{\psi(\vec{p})}^2 d^3p
\qquad\Rightarrow\qquad
\psi(\vec{v})=(4\pi\rho\Pl^2)^{3/2}
\left(\frac{\Delta^3}{p_1p_2p_3}\right)^{1/4}\psi(\vec{p}).
\ee

In $L^2(\mathbb{R}^3_\mathrm{Bohr},d^3v_\mathrm{Bohr})$, we then define the unitary operators
\ba
\hat{J}^I&:=&\widehat{e^{-2\frac{\partial}{\partial v_I}}}
\sim e^{+i\,\mubar_1 c^1}
\qquad\text{or}\quad
\hat{J}^1\psi(\vec{v})=\psi(v_1-2,v_2,v_3),\\
\hat{J}^{I\dagger}&:=&\widehat{e^{+2\frac{\partial}{\partial v_I}}}
\sim e^{-i\,\mubar_1 c^1}
\qquad\text{or}\quad
\hat{J}^{1\dagger}\psi(\vec{v})=\psi(v_1+2,v_2,v_3).
\ea
Note that $\hat{J}^I$ and $\hat{J}^{I\dagger}$ correspond to their classical counterparts $e^{+i\,\mubar_1 c^1}$ and $e^{-i\,\mubar_1 c^1}$, respectively, and we have
\be\label{eqn:commutators of v and J}
[\hat{v}_I,\hat{J}^J]=2\delta_I^J\hat{J}^J,
\qquad
[\hat{v}_I,\hat{J}^{J\dagger}]=-2\delta_I^J\hat{J}^{J\dagger},
\qquad
\hat{J}^I\hat{J}^{J\dagger}=\hat{J}^{J\dagger}\hat{J}^I=\delta^I_J.
\ee
For later use, We also define the hermitian operators
\ba
\hat{\bK}^I&:=&6\pi\gamma\Pl^2\,\frac{1}{2i}
\left[\hat{v}_I\hat{J}^I-\hat{J}^{I\dagger}\hat{v}_I\right],\\
\hat{\bS}_I&:=&6\pi\gamma\Pl^2\,\frac{1}{2}
\left[\hat{v}_I\hat{J}^I+\hat{J}^{I\dagger}\hat{v}_I\right],
\ea
which correspond to the classical counterparts:
\ba
\label{eqn:K and sin}
\hat{\bK}^I&\sim& \frac{\sin(\mubar_Ic^I)}{\mubar_I}p_I
\quad \mathop{\longrightarrow}\limits_{\mubar_Ic^I\rightarrow0}\; p_Ic^I,\\
\label{eqn:S and cos}
\hat{\bS}_I&\sim& \frac{\cos(\mubar_Ic^I)}{\mubar_I}p_I =6\pi\gamma\Pl^2 \cos(\mubar_Ic^I)v_I
\quad \mathop{\longrightarrow}\limits_{\mubar_Ic^I\rightarrow0}\; 6\pi\gamma\Pl^2 v_I=p_I^{3/2}\Delta^{-1/2}.
\ea
By \eqnref{eqn:commutators of v and J}, we have
\be\label{eqn:commutators of v K S}
[\hat{v}_I,\hat{\bK}^J]=-2i\delta_I^J\hat{\bS}_I,
\qquad
[\hat{v}_I,\hat{\bS}_J]=2i\delta_J^I\hat{\bK}^I,
\qquad
[\hat{\bK}^I,\hat{\bS}_J]=-2i(6\pi\gamma\Pl^2)^2\delta^I_J
\left(1-\hat{v}_J\right),
\ee
and obviously
\be\label{eqn:commutators of v K S 2}
[\hat{v}_I,\hat{v}_J]=
[\hat{\bK}^I,\hat{\bK}^J]=
[\hat{\bS}_I,\hat{\bS}_J]=0.
\ee

Now, in the kinematic Hilbert space to be $\hilbert_\mathrm{kin}^\mathrm{LQC}=L^2(\mathbb{R}^3_\mathrm{Bohr},d^3v_\mathrm{Bohr})\otimes L^2(\mathbb{R},d\phi)$, let the Hamiltonian given by \eqnref{eqn:H bar} promoted to the hermitian Hamiltonian operator:
\be\label{eqn:H bar hat}
\hat{\bar{H}}
=-\frac{1}{8\pi G\gamma^2}
\left(\hat{\bK}^2\hat{\bK}^3+\hat{\bK}^1\hat{\bK}^3+\hat{\bK}^1\hat{\bK}^2\right)
+\frac{\hat{p}_\phi^2}{2},
\ee
where $\hat{p}_\phi\rightarrow -i\hbar\frac{\partial}{\partial\phi}$ is the same as that of the WDW theory studied in \secref{sec:effective WDW}. The quantum Hamiltonian given by \eqnref{eqn:H bar hat} is exactly the same as that of the simplified LQC studied in \cite{Szulc:2008ar}. We will take \eqnref{eqn:H bar hat} as our fundamental quantum theory and study its effective equations of motion, which are to be compared with the numerical results obtained in \cite{Szulc:2008ar}.\footnote{Both in the detailed construction of LQC \cite{Chiou:2006qq} and in the simplified LQC \cite{Szulc:2008ar}, to extract the physical information, we have to further restrict ourselves to one of the super-selected sectors $\hilbert^{\vec \epsilon}_\mathrm{phys}$ as our \emph{physical} Hilbert space and define the physical inner product $\mean{\cdot|\cdot}^{\vec \epsilon}_\mathrm{phys}$ on it. In Approach II, as commented in \secref{sec:Approach II}, we do not need to construct the physical Hilbert space $\hilbert_\mathrm{phys}$ but still all the information about dynamics can be extracted solely from the \emph{kinematic} Hilbert space $\hilbert_\mathrm{kin}$ endowed with the original inner product $\mean{\cdot|\cdot}$.}

\subsection{Classical and quantum variables}
\label{sec:LQC variables}
Before we tackle the effective equations of motion for the simplified LQC, we have to define the classical and quantum variables and the Poisson brackets of them. By choosing $(v_I,\bK^J)$ or $(\bS_I,\bK^J)$ as the fundamental variables for the gravitational part, the classical variables are defined as
\be\label{eqn:v K S}
v_I:=\mean{\hat{v}_I},\qquad
\bK^I:=\mean{\hat{\bK}^I},\qquad
\bS_I:=\mean{\hat{\bS}_I}
\ee
in addition to \eqnref{eqn:phi pphi}.
By \eqnref{eqn:commutators of v K S} and \eqnref{eqn:commutators of v K S 2}, \eqnref{eqn:symplectic structure} then gives the Poisson brackets:
\ba\label{eqn:LQC Poisson brackets 1}
\{v_I,\bK^J\}=-\frac{2}{\hbar}\delta_I^J \bS_I,\quad
\{v_I,\bS_J\}=\frac{2}{\hbar}\delta^I_J\bK^I,\quad
\{\bK^I,\bS_J\}=\frac{2}{\hbar}(6\pi\gamma\Pl^2)^2\delta^I_J(1-v_J),
\ea
and
\be\label{eqn:LQC Poisson brackets 2}
\{v_I,v_J\}=\{\bK^I,\bK^J\}=\{\bS_I,\bS_J\}=0.
\ee
The associated quantum variables of the 2nd order are also defined:
\ba
\bG^{(n=2)}:\qquad\quad
\bG_{IJ}&:=&\mean{(\hat{\bS}_I-\mean{\hat{\bS}_I})(\hat{\bS}_J-\mean{\hat{\bS}_J})},\nn\\
\bG^{IJ}&:=&\mean{(\hat{\bK}^I-\mean{\hat{\bK}^I})(\hat{\bK}^J-\mean{\hat{\bK}^J})},\nn\\
\bG^I_J&:=&\mean{(\hat{\bK}^I-\mean{\hat{\bK}^I})
(\hat{\bS}_J-\mean{\hat{\bS}_J})}_\mathrm{Weyl},\\
\text{and}\quad
\bG_{I'J'}&:=& (6\pi\gamma\Pl^2)^2
\mean{(\hat{v}_I-\mean{\hat{v}_I})(\hat{v}_J-\mean{\hat{v}_J})},\nn\\
\bG_{IJ'}&:=& (6\pi\gamma\Pl^2)
\mean{(\hat{\bS}_I-\mean{\hat{\bS}_I})(\hat{v}_J-\mean{\hat{v}_J})}_\mathrm{Weyl},\nn\\
\bG^I_{J'}&:=& (6\pi\gamma\Pl^2)
\mean{(\hat{\bK}^I-\mean{\hat{\bK}^I})(\hat{v}_J-\mean{\hat{v}_J})}_\mathrm{Weyl}.
\ea
Here, the primed lower indices are used for $\bG^{(n=2)}$ whenever $\hat{v_I}$ is referred to; the unprimed lower indices are reserved for $\hat{\bS}_I$.

Note that $v_I$, $K^I$ and $\bS_I$ are not independent of one another but related via
\be
\mbox{$\hat{\bK}^I$}^2+\mbox{$\hat{\bS}_I$}^2
=\frac{(6\pi\gamma\Pl^2)^2}{2}\left[(\hat{v}_I+2)^2+\hat{v}_I^2\right].
\ee
We suppose that the universe does not go down to the deep Planckian regime, as the big bounces of $v_I$ are expected to takes place at a much larger scale.\footnote{As will be seen in \eqnref{eqn:v crit}, this assumption can always be achieved by increasing $\mean{p_\phi}$. This also justifies why the quantum corrections on the inverse triad due to Thiemann's trick can be ignored in the simplified LQC.}
Hence, we assume $v_I\gg1$ and then have
\be\label{eqn:K2+S2}
\mbox{$\bK^I$}^2+\mbox{$\bS_I$}^2
\approx (6\pi\gamma\Pl^2)^2 v_I^2-\bG^{II}-\bG_{II}+\bG_{I'I'}.
\ee
For the same reason, up to the error due to quantum ordering, we can approximate
\ba\label{eqn:G primed and unprimed}
\bG_{IJ} &\approx& \cos(\mubar_Ic^I)\cos(\mubar_Jc^J)\bG_{I'J'}+{\cal O}(\Pl^4),\nn\\
\bG_{IJ'} &\approx& \cos(\mubar_Ic^I)\bG_{I'J'}+{\cal O}(\Pl^4),\nn\\
\bG^I_J &\approx& \cos(\mubar_Jc^J)\bG^I_{J'}+{\cal O}(\Pl^4).
\ea

Equation \eqnref{eqn:K2+S2} gives us
\ba
\label{eqn:approx 1}
\bS_I&\approx& \pm(6\pi\gamma\Pl^2)v_I
\left[1-\frac{\mbox{$\bK^I$}^2+\bG^{II}}{(6\pi\gamma\Pl^2)^2v_I^2}
+\frac{\bG_{I'I'}-\bG_{II}}{(6\pi\gamma\Pl^2)^2v_I^2}\right]^{1/2}
\nn\\
&\approx& \pm(6\pi\gamma\Pl^2)v_I
\left[1
-\frac{\mbox{$\bK^I$}^2+\bG^{II}}{(6\pi\gamma\Pl^2)^2v_I^2}
+\frac{\mbox{$\bK^I$}^2\bG_{I'I'}}{(6\pi\gamma\Pl^2)^4v_I^4}
\right]^{1/2},
\ea
where in the second line we have used
\ba
\bG_{I'I'}-\bG_{II}&\approx&[1-\cos^2(\mubar_Ic^I)]\bG_{I'I'}
\approx\left[1-\frac{\bS^2}{(6\pi\gamma\Pl^2)^2v_I^2}\right]\bG_{I'I'}\nn\\
&\approx&\frac{\mbox{$\bK^I$}^2+{\cal O}(G^{(n=2)})}{(6\pi\gamma\Pl^2)^2v_I^2}\,\bG_{I'I'}
\ea
to trade $\bG_{II}$ for $\bG_{I'I'}$.
Equations \eqnref{eqn:approx 1} can be understood from \eqnref{eqn:S and cos}: While $\bS_I\approx (6\pi\gamma\Pl^2)v_I$ in the classical regime ($\mubar_Ic^I\rightarrow0$ and $\cos(\mubar_Ic^I)\approx 1$), in the quantum regime where the LQC corrections become significant, $\cos(\mubar_Ic^I)$ eventually crosses zero and thus $\bS_I$ flips signs ($\pm$) when $v_I$ approaches a certain critical value.
Similarly, \eqnref{eqn:G primed and unprimed} leads to
\ba
\label{eqn:approx 2a}
\bG_{IJ}
&\approx&
\left(
\pm\left[1-\frac{\mbox{$\bK^I$}^2+\bG^{II}}{(6\pi\gamma\Pl^2)^2v_I^2}
+\frac{\mbox{$\bK^I$}^2\bG_{I'I'}}{(6\pi\gamma\Pl^2)^4v_I^4}\right]^{1/2}
\right)\nn\\
&&\quad\times
\left(
\pm\left[1-\frac{\mbox{$\bK^J$}^2+\bG^{II}}{(6\pi\gamma\Pl^2)^2v_J^2}
+\frac{\mbox{$\bK^J$}^2\bG_{J'J'}}{(6\pi\gamma\Pl^2)^4v_J^4}
\right]^{1/2}
\right)
\bG_{I'J'},\\
\label{eqn:approx 2b}
\bG_{IJ'} &\approx& \pm\left[1-\frac{\mbox{$\bK^I$}^2+\bG^{II}}{(6\pi\gamma\Pl^2)^2v_I^2}
+\frac{\mbox{$\bK^I$}^2\bG_{I'I'}}{(6\pi\gamma\Pl^2)^4v_I^4}\right]^{1/2}
\bG_{I'J'},\\
\label{eqn:approx 2c}
\bG^I_J &\approx& \pm\left[1-\frac{\mbox{$\bK^J$}^2+\bG^{JJ}}{(6\pi\gamma\Pl^2)^2v_J^2}
+\frac{\mbox{$\bK^J$}^2\bG_{J'J'}}{(6\pi\gamma\Pl^2)^4v_J^4}\right]^{1/2}
\bG^I_{J'}.
\ea

As in \secref{sec:WDW variables}, the quantum variables do not commute with the classical variables, which are not chosen to be canonical variables. By \eqnref{eqn:commutators of v K S}, \eqnref{eqn:commutators of v K S 2} and \eqnref{eqn:symplectic structure}, we list the Poisson brackets needed for the later use:
\ba
&&\{v_I,\bG_{JK}\}=\frac{2}{\hbar}\left(\delta_I^J \bG_K^J+\delta_I^K \bG_J^K\right),\qquad
\{v_I,\bG^{JK}\}=-\frac{2}{\hbar}\left(\delta_I^J \bG_J^K+\delta_I^K \bG_K^J\right),\nn\\
&&\{v_I,\bG^J_K\}=-\frac{2}{\hbar}\left(\delta_I^J \bG_{JK}-\delta_I^K \bG^{JK}\right),
\ea
and
\ba
&&\{\bK^I,\bG_{JK}\}=12\pi G\gamma\left(\delta^I_J \bG_{J'K} +\delta^I_K \bG_{JK'}\right)\bG_{JK},\qquad
\{\bK^I,\bG^J_K\}=12\pi G\gamma \delta^I_K \bG^J_{K'},\nn\\
&&\{\bK^I,\bG_{J'K'}\}=12\pi G\gamma\left(\delta^I_J \bG_{JK'} +\delta^I_K \bG_{J'K}\right)\bG_{JK},\qquad
\{\bK^I,\bG^J_{K'}\}=12\pi G\gamma \delta^I_K \bG^J_K,\nn\\
&&\{\bK^I,\bG^{JK}\}=0.
\ea
Similar to \eqnref{eqn:commutators of G and G}, we also have
\be
\{\bG^{(n=2)},\bG^{(n=2)}\}=8\pi G\gamma\, \bG^{(n=3)}.
\ee

Finally, the Schwartz inequality gives the relations:
\ba
\label{eqn:LQC Schwartz 1}
\bG^{II}\bG^{JJ}&\geq& \mbox{$\bG^{IJ}$}^2,\\
\label{eqn:LQC Schwartz 2}
\bG_{II}\bG_{JJ}&\geq& \mbox{$\bG_{IJ}$}^2,\qquad\qquad
\bG_{I'I'}\bG_{J'J'}\geq \mbox{$\bG_{I'J'}$}^2\\
\label{eqn:LQC Schwartz 3}
\bG^{II}\bG_{JJ}&\geq& \mbox{$\bG^{I}_J$}^2+\frac{1}{4}\abs{\mean{[\hat{\bK}^I,\hat{\bS}_J]}}^2
=\mbox{$\bG^I_J$}^2+(6\pi\gamma\Pl^2)^4\delta^I_J(1-v_I)^2,\\
\label{eqn:LQC Schwartz 4}
\bG^{II}\bG_{J'J'}&\geq& \mbox{$\bG^{I}_{J'}$}^2+\frac{(6\pi\gamma\Pl^2)^2}{4}\abs{\mean{[\hat{\bK}^I,\hat{v}_J]}}^2
=\mbox{$\bG^I_{J'}$}^2+(6\pi\gamma\Pl^2)^2\delta^I_J \bS_I^2.
\ea

Also note that all the equations regarding $\phi$ and $p_\phi$ are exactly the same as those appeared in \secref{sec:WDW variables}.

\subsection{Effective equations of motion and Hamiltonian constraint}\label{sec:LQC eom}
Corresponding to \eqnref{eqn:H bar hat}, the quantum Hamiltonian defined in \eqnref{eqn:HQ} reads as
\be
\bar{H}_Q=-\frac{1}{8\pi G\gamma^2}\left(\bK^2\bK^3+\bK^1\bK^3+\bK^1\bK^2+\bG^{23}+\bG^{13}+\bG^{12}\right)
+\frac{p_\phi^2}{2}+\frac{G_{p_\phi p_\phi}}{2}.
\ee
According to \eqnref{eqn:Hamilton eq 1} and the Poisson brackets listed in \secref{sec:LQC variables}, the equations of motion for the classical variables are given by
\ba
\label{eqn:LQC eom1}
\frac{d\bK^I}{dt'}&=&\{\bK^I,\bar{H}_Q\}=0\quad\Rightarrow\quad \bK^I:=8\pi G\gamma\hbar\,{\calK}^I \text{ are constant},\\
\label{eqn:LQC eom2}
\frac{dv_1}{dt'}&=&\{v_1,\bar{H}_Q\}=\frac{1}{4\pi G\hbar\gamma^2}\left\{(\bK^2+\bK^3)\bS_1+\bG_1^2+\bG_1^3\right\}.
\ea
Meanwhile, $d\phi/dt'$ and $dp_\phi/dt'$ are unchanged, as given by \eqnref{eqn:wdw eom3} and \eqnref{eqn:wdw eom4}, respectively.
Similarly, by \eqnref{eqn:Hamilton eq 2}, the quantum variables satisfy the following equations of motion:
\ba
\label{eqn:LQC eom3}
\frac{d\bG^{IJ}}{dt'}&=&\{\bG^{IJ},\bar{H}_Q\}=0\quad\Rightarrow\quad \bG^{IJ}=g^{IJ} \text{ are constant},\\
\label{eqn:LQC eom4}
\frac{d\bG_{1'2'}}{dt'}&=&\{\bG_{1'2'},\bar{H}_Q\}
=\frac{3}{2}\gamma^{-1}\left\{\bG_{12'}
\left(\bK^2+\bK^3\right)+\bG_{21'}\left(\bK^1+\bK^3\right)\right\}\nn\\
&&\qquad\qquad\qquad\quad+\gamma^{-1}{\cal O}(\bG^{(n=3)}),\\
\label{eqn:LQC eom5}
\frac{d\bG_{1'1'}}{dt'}&=&\{\bG_{1'1'},\bar{H}_Q\}
=\frac{3}{2}\gamma^{-1}\bG_{11'}\left(2\bK^2+2\bK^3\right)+\gamma^{-1}{\cal O}(\bG^{(n=3)}),\\
\label{eqn:LQC eom6}
\frac{d\bG^I_{1'}}{dt'}&=&\{\bG^I_{1'},\bar{H}_Q\}
=\frac{3}{2}\gamma^{-1}\bG^I_1\left(\bK^2+\bK^3\right)+\gamma^{-1}{\cal O}(\bG^{(n=3)}),
\ea
in addition to \eqnref{eqn:wdw eom9}, \eqnref{eqn:wdw eom10} and \eqnref{eqn:wdw eom11}.

The right-hand side of the above equations of motion involves $\bS_I$ and $\bG^{(n=2)}$ with unprimed lower indices, although we consider only the time derivative of $v_I$ and $\bG^{(n=2)}$ with primed lower indices on the left-hand side. Using \eqnref{eqn:approx 1} and \eqnref{eqn:approx 2a}--\eqnref{eqn:approx 2c} to eliminate $\bS_I$ and $\bG^{(n=2)}$ with unprimed lower indices in favor of $v_I$ and $\bG^{(n=2)}$ with only primed lower indices, we get
\ba
\label{eqn:LQC eom7}
\bK^I&=&8\pi G\gamma\hbar\,{\calK}^I \text{ are constant},\\
\label{eqn:LQC eom8}
\bG^{IJ}&=&g^{IJ} \text{ are constant},\\
\label{eqn:LQC eom9}
p_\phi\frac{dv_1}{d\phi}&\approx&\pm\frac{3}{2}\gamma^{-1}v_1
\left[1-\frac{\mbox{$\bK^1$}^2+\bG^{11}}{(6\pi\gamma\Pl^2)^2v_1^2}
+\frac{\mbox{$\bK^1$}^2\bG_{1'1'}}{(6\pi\gamma\Pl^2)^4v_1^4}
\right]^{1/2}\nn\\
&&\qquad\quad\times
\left\{
\left(\bK^2+\bK^3\right)+\frac{\bG^2_{1'}+\bG^3_{1'}}{6\pi\gamma\Pl^2v_1}
\right\},
\ea
and
\ba
\label{eqn:LQC eom10}
p_\phi\frac{d\bG_{1'2'}}{d\phi}&\approx&
\frac{3}{2}\gamma^{-1}\bG_{1'2'}
\Biggl\{
\pm\left[1-\frac{\mbox{$\bK^1$}^2+\bG^{11}}{(6\pi\gamma\Pl^2)^2v_1^2}
+\frac{\mbox{$\bK^1$}^2\bG_{1'1'}}{(6\pi\gamma\Pl^2)^4v_1^4}
\right]^{1/2}
\left(\bK^2+\bK^3\right)\nn\\
&&\qquad\qquad\quad
\pm\left[1-\frac{\mbox{$\bK^2$}^2+\bG^{22}}{(6\pi\gamma\Pl^2)^2v_2^2}
+\frac{\mbox{$\bK^2$}^2\bG_{2'2'}}{(6\pi\gamma\Pl^2)^4v_2^4}
\right]^{1/2}
\left(\bK^1+\bK^3\right)
\Biggr\},\\
\label{eqn:LQC eom11}
p_\phi\frac{d\bG_{1'1'}}{d\phi}&\approx&
\pm\frac{3}{2}\gamma^{-1}\bG_{1'1'}
\left[1-\frac{\mbox{$\bK^1$}^2+\bG^{11}}{(6\pi\gamma\Pl^2)^2v_1^2}
+\frac{\mbox{$\bK^1$}^2\bG_{1'1'}}{(6\pi\gamma\Pl^2)^4v_1^4}
\right]^{1/2}
\left(2\bK^2+2\bK^3\right),\\
\label{eqn:LQC eom12}
p_\phi\frac{d\bG^I_{1'}}{d\phi}&\approx&
\pm\frac{3}{2}\gamma^{-1}\bG^I_{1'}
\left[1-\frac{\mbox{$\bK^1$}^2+\bG^{11}}{(6\pi\gamma\Pl^2)^2v_1^2}
+\frac{\mbox{$\bK^1$}^2\bG_{1'1'}}{(6\pi\gamma\Pl^2)^4v_1^4}
\right]^{1/2}
\left(\bK^2+\bK^3\right),
\ea
where we have used \eqnref{eqn:wdw eom3} and treat $\phi$ as the internal time.

When $v_I$ approaches a certain critical value, the factor $\pm[-\cdots]^{1/2}$ flips signs and consequently, by \eqnref{eqn:LQC eom9}, $v_I$ undergoes the bouncing behavior at the critical point, affirming the bouncing scenario predicted in \cite{Chiou:2006qq,Chiou:2007dn,Chiou:2007mg}.

If the off-diagonal squeezing is negligible (i.e. $\bG^I_{J'}/(6\pi\gamma\Pl^2\bK^I v_J)\ll 1$ for $I\neq J$), the second term in the curly bracket of \eqnref{eqn:LQC eom9} can be dropped. Equations \eqnref{eqn:LQC eom9} and \eqnref{eqn:LQC eom11} then yield
\be
\frac{d\bG_{I'I'}}{dv_I}\approx 2\frac{\bG_{I'I'}}{v_I}
\qquad\Rightarrow\qquad
\eta^2_{v_I}
\approx \text{ constant},
\ee
where the dimensionless quantity
\be
\eta^2_{v_I}:=\frac{\Delta^2v_I}{v_I^2}
=\frac{\bG_{I'I'}}{(6\pi\gamma\Pl^2)^2v_I^2}
\ee
is the relative spread of $v_I$.
In the case without off-diagonal squeezing, the critical value of $v_I$ is given by the constants $\bK^I$, $\bG^{II}$ and $\eta^2_{v_I}$ as
\be\label{eqn:v crit}
v^2_{I, \mathrm{crit}}=\frac{\mbox{$\bK^I$}^2
\left(1+\eta^2_{v_I}+\eta^2_{\bK^I}\right)}
{(6\pi\gamma\Pl^2)^2},
\ee
where the dimensionless \emph{constant}
\be
\eta^2_{\bK^I}:=\frac{\Delta^2\bK^I}{\mbox{$\bK^I$}^2}
=\frac{\bG^{II}}{\mbox{$\bK^I$}^2}
\ee
is the relative spread of $\bK^I$.
In terms of $p_I$ via \eqnref{eqn:p and v}, \eqnref{eqn:LQC eom9} can be recast as
\be\label{eqn:LQC dp/dphi}
\frac{1}{p_I}\frac{dp_I}{d\phi}\approx
\pm\sqrt{8\pi G}\,\left(\frac{\calK^2+\calK^3}{\calK_\phi}\right)
\left[1-\frac{\varrho_I}{\varrho_{I\!,\,\mathrm{crit}}}\right]^{1/2},
\ee
where we define the \emph{directional density} for the direction $I$ as
\be
\varrho_I:=\frac{p_\phi^2}{p_I^3}
\ee
and its critical value is given by
\be\label{eqn:varrho}
\varrho_{I\!,\,\mathrm{crit}}:=\left(\frac{\calK_\phi}{\calK_I}\right)^2
\left(1+\eta_{v_I}^2+\eta_{\bK^I}^2\right)^{-1}
\rho_\mathrm{Pl}
\ee
with
\be
\qquad
\rho_\mathrm{Pl}:=(8\pi G \gamma^2\Delta)^{-1}
\ee
being the \emph{Planckian density}.
It tells that the big bounces occur up to three times, once in each direction of $p_I$, whenever each of the directional density approaches its critical value and thus flips the sign of \eqnref{eqn:LQC dp/dphi}.
This closely agrees with the semiclassical description in \cite{Chiou:2007dn} and \cite{Chiou:2007mg} except that the critical value $\varrho_{I,\mathrm{crit}}$ is now slightly modified by the overall factor $(1+\eta_{v_I}^2+\eta_{\bK^I}^2)$ involving the relative spreads and giving rise to the back-reaction. The back-reaction due to uncertainties in $\bK^I$ and $v_I$ is expected and desired, because one would otherwise be puzzled why the critical value $\varrho_{I,\mathrm{crit}}$ is exactly determined by the ratio of $\calK_I$ and $\calK_\phi$, which are the parameters describing the classical behavior and should not completely dictate the quantum physics of the bouncing scenario.

In the case with appreciable off-diagonal squeezing ($\bG_{I'}^J\not\ll 6\pi\gamma\Pl^2v_I\bK^J$ for $I\neq J$), the terms involving $\bG_{I'}^J$ in \eqnref{eqn:LQC eom9} will further deviate the trajectory of $v_I$ and the relative spreads $\eta_{v_I}^2$ will no longer stay constant. Furthermore, the evolution of $v_I$ in different diagonal directions couples to one another through the off-diagonal squeezing. Also note that the second term inside the curly bracket in \eqnref{eqn:LQC eom9} is $\propto v_I^{-1}$ and therefore this back-reaction due to off-diagonal squeezing is most significant in the vicinity of the big bounce and negligible in the classical regime.

Finally, the Hamiltonian constraint is exactly the same as that studied in \secref{sec:Hamiltonian constraint} except that all $\hat{K}^I$, $K^I$ and $G^{IJ}$ should be replaced by $\hat{\bK}^I$, $\bK^I$ and $\bG^{IJ}$. The validity of treating $\phi$ as the internal time follows the same argument given in \secref{sec:WDW eom}.

\section{Symmetry reduction to the isotropic model}\label{sec:symmetry reduction}
One of the most challenging problems in LQC is to find a systematic procedure to derive LQC as a symmetry-reduced theory from the full theory of LQG. To shed some light on this issue, we can ask a similar but much easier question: ``How does isotropy emerge from the anisotropic Bianchi I description?'' With the effective theory at hand, we try to answer this question and make sense of the isotropy reduction from the Bianchi I model. We will discuss two ways of reduction: the \emph{formal} and \emph{physical} prescriptions.

In the framework of effective equations of motion, similar to the Bianchi I model, the $k=0$ FRW model has the classical variables:
\be
v:=\mean{\hat{v}},\qquad
\bK:=\mean{\hat{\bK}},\qquad
\bS:=\mean{\hat{\bS}},\qquad
\ee
and the associated quantum variables are given by
\ba
\bG^{(n=2)}:\qquad\quad
\bG_{(2)}&:=&\mean{(\hat{\bS}-\mean{\hat{\bS}})(\hat{\bS}-\mean{\hat{\bS}})},\nn\\
\bG^{(2)}&:=&\mean{(\hat{\bK}-\mean{\hat{\bK}})(\hat{\bK}-\mean{\hat{\bK}})},\nn\\
\bG^{(1)}_{(1)}&:=&\mean{(\hat{\bK}-\mean{\hat{\bK}})(\hat{\bS}-\mean{\hat{\bS}})}_\mathrm{Weyl},\\
\text{and}\quad
\bG_{(2'')}&:=& (6\pi\gamma\Pl^2)^2
\mean{(\hat{v}-\mean{\hat{v}})(\hat{v}-\mean{\hat{v}})},\nn\\
\bG_{(11')}&:=& (6\pi\gamma\Pl^2)
\mean{(\hat{\bS}-\mean{\hat{\bS}})(\hat{v}-\mean{\hat{v}})}_\mathrm{Weyl},\nn\\
\bG^{(1)}_{(1')}&:=& (6\pi\gamma\Pl^2)
\mean{(\hat{\bK}-\mean{\hat{\bK}})(\hat{v}-\mean{\hat{v}})}_\mathrm{Weyl}.
\ea

To get the effective equations of motion for the isotropic case from those for the Bianchi I model, we can \emph{formally} impose the condition on the Bianchi I variables to demand that the classical and quantum variables with different indices are all identical; i.e.,
\ba\label{eqn:isotropy 1}
&&\quad v_1=v_2=v_3\equiv v,\quad\text{(i.e. }p_1=p_2=p_3\equiv p\text{)}, \nn\\
&&\quad \bK^1=\bK^2=\bK^3\equiv \bK,\qquad \bS_1=\bS_2=\bS_3\equiv \bS,\nn\\
&&\bG_{11}=\bG_{22}=\bG_{33}=\bG_{12}=\bG_{13}=\bG_{23}\equiv \bG_{(2)}
\quad\text{and so on for other $\bG^{(n=2)}$},
\ea
which is trivially compatible with \eqnref{eqn:LQC eom1}--\eqnref{eqn:LQC eom12}. With this condition imposed, the equations of motion reduce to the isotropic counterparts. In particular, \eqnref{eqn:LQC eom9} becomes
\be\label{eqn:iso sol 1}
p_\phi\frac{dv}{d\phi}\approx\pm 3\gamma^{-1}v
\left[1-\frac{(\bK)^2+\bG^{(2)}}{(6\pi\gamma\Pl^2)^2v^2}
+\frac{(\bK)^2\,\bG_{(2')}}{(6\pi\gamma\Pl^2)^4v^4}
\right]^{1/2}
\left\{
\bK+\frac{\bG^{(1)}_{(1')}}{6\pi\gamma\Pl^2v}
\right\},
\ee
where
\be\label{eqn:iso K}
\bK=8\pi G\gamma\hbar\calK
\qquad\text{and}\qquad
p_\phi=\hbar\sqrt{8\pi G}\calK_\phi,
\ee
and the Hamiltonian constraint \eqnref{eqn:K} now reads as
\be\label{eqn:iso K and Kphi}
\calK_\phi^2=6\calK^2.
\ee
Equations \eqnref{eqn:iso sol 1} can be recast as [cf. \eqnref{eqn:LQC dp/dphi}--\eqnref{eqn:varrho}]
\be\label{eqn:iso dp/dphi}
\frac{1}{dp}\frac{dp}{d\phi}\approx
\pm\sqrt{\frac{16\pi G}{3}}\left[1-\frac{\rho_\phi}{\rho_\mathrm{crit}}\right]^{1/2}
\left\{
1+\frac{\bG^{(1)}_{(1')}}{6\pi\gamma\Pl^2\bK v}
\right\},
\ee
where $\rho_\phi$ is the matter density of $\phi$:
\be
\rho_\phi:=\frac{p_\phi^2}{2p^3}
\ee
and its critical value is given by
\be\label{eqn:iso rho crit}
\rho_\mathrm{crit}=3\left(1+\eta_{v}^2+\eta_{\bK}^2\right)^{-1}
\rho_\mathrm{Pl}< 3\rho_\mathrm{Pl}.
\ee
This perfectly concurs with the exact solution obtained in \cite{Ashtekar:2006es}, which indicates that the matter density has an absolute upper bound and the more semiclassical the state is (correspondingly, $\eta_{v}^2$ and $\eta_{\bK}^2$ are smaller), the closer the upper bound can be reached at the big bounce.

Although this procedure formally gives rise to the isotropic theory, in the context of the Bianchi I model, however, the solution satisfying condition \eqnref{eqn:isotropy 1} is \emph{not} ``isotropic'' at all. In fact, it is the most anisotropic one in the sense that three different directions are now completely correlated, leaving the spatial direction aligned with the vector $(x=1,y=1,z=1)$ distinct from any other directions. To sensibly describe an isotropic solution \emph{within} the Bianchi I formulation, we should have no off-diagonal correlations; that is, instead of the \emph{formal} isotropic condition \eqnref{eqn:isotropy 1}, we should impose the \emph{physical} isotropic condition:
\ba\label{eqn:isotropy 2}
&&v_1=v_2=v_3\equiv v,\quad
\bK^1=\bK^2=\bK^3\equiv \bK,\qquad
\bS_1=\bS_2=\bS_3\equiv \bS,\nn\\
&&\bG_{11}=\bG_{22}=\bG_{33}\equiv \bG_{(2)}
\quad\text{but}\quad
\bG_{12}=\bG_{13}=\bG_{23}=0,\nn\\
&&\ \text{and similarly for other $\bG^{(n=2)}$}.
\ea
It is nontrivial to note that this prescription is also compatible with \eqnref{eqn:LQC eom1}--\eqnref{eqn:LQC eom12}. Particularly, the physical isotropic condition reduces \eqnref{eqn:LQC eom9} to
\be\label{eqn:iso sol 2}
p_\phi\frac{dv}{d\phi}\approx\pm 3\gamma^{-1} v
\left[1-\frac{(\bK)^2+\bG^{(2)}}{(6\pi\gamma\Pl^2)^2v^2}
+\frac{(\bK)^2\,\bG_{(2')}}{(6\pi\gamma\Pl^2)^4v^4}
\right]^{1/2}(\bK)^2
\ee
and \eqnref{eqn:K} to
\be
\calK_\phi^2=6\calK\left(1-\frac{\eta^2_\bK}{3}\right).
\ee
Consequently, we have [cf. \eqnref{eqn:iso dp/dphi}]
\be
\frac{1}{dp}\frac{dp}{d\phi}\approx
\pm\sqrt{\frac{16\pi G}{3}}\left[1-\frac{\rho_\phi}{\rho_\mathrm{crit}}\right]^{1/2},
\ee
with the critical value of $\rho_\phi$ given by [cf. \eqnref{eqn:iso rho crit}]
\be\label{eqn:iso rho crit 2}
\rho_\mathrm{crit}=
3\left(\frac{1-\eta^2_\bK/3}{1+\eta_{v}^2+\eta_{\bK}^2}\right)
\rho_\mathrm{Pl}< 3\rho_\mathrm{Pl}.
\ee

The physical prescription \eqnref{eqn:isotropy 2} yields the effective equations slightly different from those by the formal prescription \eqnref{eqn:isotropy 1}. Comparing \eqnref{eqn:iso sol 2}, (which gives isotropic solutions within the Bianchi I model) with \eqnref{eqn:iso sol 1}, (which describes the solution in the isotropic model), we note that, at least up to $\bG^{(n=2)}$, the equation of motion for $p$ subject to \eqnref{eqn:isotropy 2} receives no back-reaction due to the squeezing $G^{(1)}_{(1')}$, while that subject to \eqnref{eqn:isotropy 1} does. Furthermore, \eqnref{eqn:iso rho crit} and \eqnref{eqn:iso rho crit 2} do not agrees precisely. The moral we learned is: In the context of effective theory, it is possible to have a well-posed solution exhibiting a certain symmetry (e.g. isotropy) within the framework which does not assume such a symmetry; however, the well-posed solution in the less symmetric (e.g. Bianchi I) theory and the solution obtained directly from the more symmetric (e.g. isotropic) theory agree only approximately.

\section{Summary and Discussion}\label{sec:discussion}
When the effective theory developed in \cite{Bojowald:2005cw} is applied to the cosmological models more complicated than the $k=0$ FRW case, we face the problem that, in the scheme of Approach I, the Hamiltonian operator is not polynomial of the fundamental operators and, as a consequence, the approximation scheme is out of good control. To work around the technical difficulties, we devise a new approach --- Approach II --- for the constrained quantum system that possesses an internal clock. In Approach II, the quantum evolution are approximated by a finite system of coupled but ordinary differential equations \eqnref{eqn:Hamilton eq 1} and \eqnref{eqn:Hamilton eq 2} for the classical and quantum variables and, on top of them, the Hamiltonian constraint is weakly imposed by \eqnref{eqn:Approach II constraint}. The evolution equations \eqnref{eqn:Hamilton eq 1} and \eqnref{eqn:Hamilton eq 2} adhered to \eqnref{eqn:Approach II constraint} can be understood as the natural extension of the classical Hamilton's equations and the classical Hamiltonian constraint.

It is tantalizing that Approach II, albeit only heuristically motivated, could be used as a viable formulation for constrained quantum systems even in the fundamental level, as it is shown to be self-consistent. Approach II has the virtue that the technical and conceptual difficulties in the standard treatment for the constrained quantum systems are avoided and the quantum evolution is posed in a very intuitive picture. Furthermore, the philosophy of timeless formulation is better actualized, as the notion of measurement of time is retained and the time variable is on the equal footing as other observables, giving insight into the problem of time in the relativistic quantum mechanics.

When applied to the WDW theory in the Bianchi I model, up to the 2nd order, Approach II gives the effective solutions \eqnref{eqn:sol for GIJ}--\eqnref{eqn:sol for pI} subject to the constraints \eqnref{eqn:K} and \eqnref{eqn:K 2}, which agree with the results obtained in the fully developed quantum theory in \cite{Chiou:2006qq} for the case without off-diagonal squeezing. We also expect that the evolution in the fully developed quantum theory receives back-reaction arising from the off-diagonal squeezing.

For simplified LQC, Approach II leads to the effective equations of motion \eqnref{eqn:LQC eom1}--\eqnref{eqn:LQC eom6} subject to the constraints \eqnref{eqn:K} and \eqnref{eqn:K 2}. The big bounces take place up to three times, once in each direction of $p_I$, whenever \eqnref{eqn:LQC eom9} flips signs. The bouncing scenario predicted in the semiclassical approach in \cite{Chiou:2007dn,Chiou:2007mg} is affirmed as the directional densities $\varrho_I$ indicate the occurrence of bounces, but the critical values $\varrho_{I,\mathrm{crit}}$ are modified by the quantum corrections as shown in \eqnref{eqn:varrho}. The off-diagonal squeezing gives further back-reaction and makes the evolution of $p_I$ in different diagonal directions couple to one another. To see if these results agree with the standard treatment of the simplified LQC, it is necessary to explore more details of the numerical computation in \cite{Szulc:2008ar}.

Additionally, the framework of effective equations offers a language to describe the isotropy-reduced solution in the context of the anisotropic Bianchi I model. The fact that the physical prescription for the isotropic condition is compatible with the effective equations of motions implies that, within the anisotropic framework, it is possible to have a well-posed solution exhibiting isotropy, which, however, is slightly different from the solution in the isotropic framework. The lesson may teach us how a certain symmetry emerges from a less symmetric model and whether a symmetric theory can be systematically derived from a more fundamental (namely, less symmetric) theory.

However, we should keep the caveat in mind that the methodology of Approach II has not been stringently validated and it is still unclear how well and why Approach II agrees with the standard treatment for constrained quantum systems for generic cases. Therefore, a more rigorous formulation directly derived from the standard quantum theory is on demand and currently under development \cite{Bojowald:2008}. Meanwhile, to test robustness of ramifications of Approach II, it will be very instructive to repeat the analysis of this paper to other extended models, such as LQC in the $k=\pm$ FRW models \cite{Ashtekar:2006es,Vandersloot:2006ws}, LQC in Kantowski-Sachs spacetime \cite{Chiou:2008eg} and the loop quantum geometry of the Schwarzschild black hole interior \cite{Chiou:2008nm}.\footnote{In the Schwarzschild black hole interior, the classical Hamiltonian constraint can be rescaled as $H=-\left(2bcp_bp_c+(b^2+\gamma^2)p_b^2\right)$. Even though there is no matter content, $p_b$ could be locally treated as the internal clock by identifying the quadratic term $\gamma^2p_b^2$ in $H$ as $p_\phi^2$.}

Finally, as commented in \secref{sec:LQC simplified LQC}, the $\mubar'$-scheme quantization has a better scaling property but the quantum theory based on it is very difficult to construct. The difficulty is due to the fact that the corresponding affine variables do not exist. (See Appendix B of \cite{Chiou:2007mg} for more details.) In the language of Approach II, the difficulty is translated into the problem that we cannot find a \emph{finite} set of fundamental variables which form a closed algebra of Poisson brackets, as opposed to \eqnref{eqn:v K S}, \eqnref{eqn:LQC Poisson brackets 1} and \eqnref{eqn:LQC Poisson brackets 2} for the $\mubar$ scheme. Nevertheless, there might be a physical sense of approximation to truncate the infinite set of fundamental variables and then the treatment of Approach II can be carried over again. This strategy might gives a sound machinery to study the effective dynamics of the $\mubar'$-scheme LQC, even if the detailed construction of the quantum theory remains unknown.

\begin{acknowledgements}
The author would like to thank Martin Bojowald for suggesting this research topic, having useful discussions and giving valuable comments. This work was supported in part by the NSF Grant PHY-0456913 and the Eberly Research Funds of The Pennsylvania State University.
\end{acknowledgements}

\appendix

\section{The WDW theory in Approach I}\label{app:WDW in Approach I}
As a comparison to Approach II in \secref{sec:effective WDW}, the effective theory of Approach I for the WDW theory is presented here along the line of \secref{sec:Approach I}.

In Approach I, the Hilbert space of the WDW theory is
$\hilbert^\mathrm{WDW}=L^2(\mathbb{R}^3,d^3p)$ and $\phi$ is treated as a pure time variable. That is, $\ket{\Psi(\phi)}\in\hilbert^\mathrm{WDW}$ is a state at the instant $\phi$ and the dynamics is govern by the Schr\"{o}dinger equation
\be
-i\hbar\frac{\partial}{\partial\phi}\ket{\Psi(\phi)}=\pm\hbar\sqrt{\hat{\Theta}}\,\ket{\Psi(\phi)}
\ee
as stated in \eqnref{eqn:Approach I constraint} with the Hamiltonian operator given by
\be\label{eqn:sqrt Theta}
\pm\hbar\sqrt{\hat{\Theta}}
=\pm\frac{1}{\sqrt{4\pi G\gamma^2}}
\left(\hat{K}^2\hat{K}^3+\hat{K}^1\hat{K}^3+\hat{K}^1\hat{K}^2\right)^{1/2}
\ee
as $\sqrt{\hat{\Theta}}$ is defined in \eqnref{eqn:H hat}.

Associated with the Hamiltonian operator in \eqnref{eqn:sqrt Theta}, the quantum Hamiltonian defined in \eqnref{eqn:HQ} yields
\be
H_Q\equiv\pm\mean{\hat{p}_\phi}
\ee
with $\mean{\hat{p}_\phi}$ given by \eqnref{eqn:constraint 2} in terms of classical and quantum variables.

Equation \eqnref{eqn:flow 1} gives the effective equations of motion:
\ba
\frac{dK^I}{d\phi}&=&\{K^I,H_Q\}=0\quad\Rightarrow\quad K^I:=8\pi G\gamma\hbar\,{\calK}^I \text{ are constant},\\
\label{eqn:dp over dphi}
\frac{dp_1}{d\phi}&=&\{p_1,H_Q\}
\approx\mp\Biggl\{\gamma^{-1}\frac{K^2+K^3}{p_{\phi,1}}\,p_1\\
&&\qquad-\sqrt{16\pi G}\left(
\frac{(K^2+K^3)^2}{4\,\theta(K^I)^{3/2}}\,G^1_1
+\frac{\theta(K^I)-{K^2}^2}{2\,\theta(K^I)^{3/2}}\,G^3_1
+\frac{\theta(K^I)-{K^3}^2}{2\,\theta(K^I)^{3/2}}\,G^2_1\right)\Biggr\},\nn
\ea
where $p_{\phi,1}^{-1}$ is short for
\ba
p_{\phi,1}^{-1}&:=&\sqrt{16\pi G\gamma^2}
\Biggl\{
\frac{1}{2\,\theta(K^I)^{1/2}}
+\frac{3(K^2+K^3)^2}{16\,\theta(K^I)^{5/2}}\,G^{11}
-\frac{(\theta(K^I)-3{K^3}^2)(K^1+K^3)}{16\,\theta(K^I)^{5/2}(K^2+K^3)}\,G^{22}\nn\\
&&\quad
-\frac{(\theta(K^I)-3{K^2}^2)(K^1+K^2)}{16\,\theta(K^I)^{5/2}(K^2+K^3)}\,G^{33}\nn\\
&&\quad
-\frac{(K^2+K^3)({K^1}^2+K^2K^3)+K^1({K^2}^2+6K^2K^3+{K^3}^2)}
{8\,\theta(K^I)^{5/2}(K^2+K^3)}\,G^{23}\nn\\
&&\quad
-\frac{\theta(K^I)-3{K^2}^2}{8\,\theta(K^I)^{5/2}}\,G^{13}
-\frac{\theta(K^I)-3{K^3}^2}{8\,\theta(K^I)^{5/2}}\,G^{12}
\Biggr\}
=p_\phi^{-1}\left[1+{\cal O}\left(\frac{G^{IJ}}{G\gamma^2p_\phi^2}\right)\right]
\ea
and so on for $p_{\phi,2}^{-1}$, $p_{\phi,3}^{-1}$ in the cyclic manner.
Similarly, by \eqnref{eqn:flow 2}, we have
\ba
\frac{dG^{IJ}}{d\phi}&=&\{G^{IJ},H_Q\}=0\quad\Rightarrow\quad G^{IJ} \text{ are constant},\\
\frac{dG_{12}}{d\phi}&=&\{G_{12},H_Q\}=\mp\gamma^{-1}G_{12}
\left\{\frac{K^2+K^3}{p_{\phi,1}}+\frac{K^1+K^3}{p_{\phi,2}}\right\}
+\gamma^{-1}p_\phi^{-1}{\cal O}(G^{(n=3)}),\\
\frac{dG_{11}}{d\phi}&=&\{G_{11},H_Q\}=\mp\gamma^{-1}G_{11}
\left\{\frac{2(K^2+K^3)}{p_{\phi,1}}\right\}+\gamma^{-1}p_\phi^{-1}{\cal O}(G^{(n=3)}),\\
\frac{dG^I_1}{d\phi}&=&\{G^I_1,H_Q\}=\mp\gamma^{-1}G^I_1
\left\{\frac{K^2+K^3}{p_{\phi,1}}\right\}+\gamma^{-1}p_\phi^{-1}{\cal O}(G^{(n=3)}).
\ea
Note that $p_{\phi,\,I}$ are constants of motion and
$p_{\phi,\,I}\approx p_\phi\equiv\mean{\hat{p}_\phi}$
if $\abs{G^{IJ}}\ll8\pi G p_\phi^2$.

The resulting effective equations of motion are qualitatively different from those in \eqnref{eqn:wdw eom1}, \eqnref{eqn:wdw eom2} and \eqnref{eqn:wdw eom5}--\eqnref{eqn:wdw eom8}. Approach II agrees with Approach I only if all the 2nd order quantum variables are ignored. In particular, \eqnref{eqn:wdw eom2} receives back-reaction only if off-diagonal squeezing are present ($G^I_J\neq0$ for $I\neq J$) but \eqnref{eqn:dp over dphi} receives back-reaction both from diagonal and off-diagonal squeezing. Furthermore, even with squeezing terms all vanishing, \eqnref{eqn:dp over dphi} still receives back-reaction through $p_{\phi,I}$ and gives a trajectory of $\mean{p_I}$ slightly deviated from the classical one, which disagrees with the result of \cite{Chiou:2006qq} for the fully developed WDW theory.

The comparison suggests that, for the Bianchi I model, the effective description of Approach I is more sensible than that of Approach II. The clumsiness of Approach II is due to the non-polynomiality of \eqnref{eqn:sqrt Theta}, which spoils the order-by-order approximation scheme. If we formally reduce the Bianchi I model to the isotropic one by taking \eqnref{eqn:isotropy 1}, it can be easily shown that $p_{\phi,1}=p_{\phi,2}=p_{\phi,3}=\mean{\hat{p}_\phi}$ and Approach II yields the same result as Approach I. That said, Approach I is good for the $k=0$ FRW model as the polynomiality of the Hamiltonian operator is recovered.

\section{Self-consistency of Approach II}\label{app:self-consistency}
In order to be a viable formulation for constrained quantum systems, Approach II has to pass two tests for self-consistency as mentioned \secref{sec:Approach II}.

For the first test, to see if the weakly imposed constraint \eqnref{eqn:Approach II constraint} still holds after the given initial state being evolved, we define the operators
\be
\hat{\Omega}_k:=(\pm\hat{p}_\phi)^k-(\hbar^2\hat{\Theta})^{k/2}
\equiv(\pm\hat{p}_\phi)^k-(-2\hat{H}_0)^{k/2}
\ee
and the associated variables
\be
\Omega_k:=\mean{\hat{\Omega}_k}
=(\pm p_\phi)^k-(-2H_0)^{k/2}
\ee
to see if $\Omega_k=0$ remains.

To check this, we compute
\ba\label{eqn:first test}
\frac{d\Omega_k}{dt'}
&=&\{\Omega_k,H_Q\}=\frac{1}{i\hbar}\mean{[\hat{\Omega}_k,\hat{H}]}\nn\\
&=&\frac{1}{2i\hbar}
\mean{[(\pm\hat{p}_\phi)^k-(-2\hat{H}_0)^{k/2},\hat{p}_\phi^2+2\hat{H}_0]}\nn\\
&=&\pm k\,\mean{(\pm\hat{p}_\phi)\,(-2\hat{H}_0)^{k/2-1}\,V'(\hat{\phi})}_\mathrm{Weyl}
\mp k\,\mean{(\pm\hat{p}_\phi)^{k-1}\,V'(\hat{\phi})}_\mathrm{Weyl},
\ea
where $V(\hat{\phi})\in\hat{H}_0$ represents the potential of $\phi$. In the case with a free massless scalar field, $V(\phi)=0$ and we have $d\Omega_k/dt'=0$ \emph{exactly}. If $V(\phi)\neq0$, the right-hand side of \eqnref{eqn:first test} does not vanish except for the case of $k=2$. Nevertheless, since we are only interested in the quantum evolution in the realm close to classical behavior, we can make the ``factorization approximation'':
\be
\mean{\hat{A}_1\hat{A}_2\cdots\hat{A}_n}_\mathrm{Weyl}\approx
\mean{\hat{A}_1}\mean{\hat{A}_2}\cdots\mean{\hat{A}_n}
\ee
for hermitian operators as long as the quantum state is semiclassical enough. Consequently, \eqnref{eqn:first test} can be approximated and gives
\be
\frac{d\Omega_k}{dt'}\approx
\pm k\left\{
\mean{\pm\hat{p}_\phi}\mean{(-2\hat{H}_0)^{k/2-1}}\mean{V'(\hat{\phi})}
-\mean{(\pm\hat{p}_\phi)^{k-1}}\mean{V'(\hat{\phi})}
\right\}
\approx 0
\ee
with the help of \eqnref{eqn:Approach II constraint}. Therefore, the first test is rigorously verified if $V(\phi)=0$ and reasonably justified if $V(\phi)\neq0$. In particular, we can see why a free massless scalar field is special for being the internal clock.

Next, for the second test, we consider the lapse function $N=N(q^i,p_i)$ associated with the time coordinate $t'$ via $d\tau=Ndt'$ with $\tau$ being the proper time. Let $H^\tau$ be the (unscaled) Hamiltonian associated with $\tau$; the rescaled Hamiltonian associated with $t'$ is given by $H^{t'}=NH^\tau$. In quantum theory, the Hamiltonian operator corresponding to $H^{t'}$ is $\hat{H}^{t'}=(\hat{N}\hat{H}^\tau)_\mathrm{Wely}$ with $\hat{N}=N(\hat{q}^i,\hat{p}_i)_\mathrm{Wely}$. With respect to $t'$, the time derivative of an arbitrary variable $\omega$ is given by
\ba
\frac{d\omega}{dt'}
&=&\{\omega,H^{t'}_Q\}=\frac{1}{i\hbar}\mean{[\hat{\omega},\hat{H}^{t'}]}
=\frac{1}{i\hbar}\mean{[\hat{\omega},(\hat{N}\hat{H}^\tau)_\mathrm{Wely}]}\nn\\
&=&\Mean{\hat{N}\left(\frac{1}{i\hbar}[\hat{\omega},\hat{H}^\tau]\right)}_\mathrm{Weyl}
+\Mean{\hat{H}^\tau\left(\frac{1}{i\hbar}[\hat{\omega},\hat{N}]\right)}_\mathrm{Weyl}\nn\\
&\approx&\mean{\hat{N}}\frac{1}{i\hbar}\mean{[\hat{\omega},\hat{H}^\tau]}
+\mean{\hat{H}^\tau}\frac{1}{i\hbar}\mean{[\hat{\omega},\hat{N}]}\nn\\
&=&\mean{\hat{N}}\{\omega,H^\tau_Q\},
\ea
where again the factorization approximation is made and we also use $\mean{\hat{H}^\tau}=0$ by \eqnref{eqn:Approach II constraint}. In particular, we have
\be
\frac{d\phi}{dt'}\approx\mean{\hat{N}}\{\phi,H^\tau_Q\}
\ee
and subsequently the factor $\mean{\hat{N}}$ cancels when we compute $d\omega/d\phi$. Therefore, the resulting effective dynamics, when related to the internal time $\phi$, is independent of different choice of the lapse function . However, it is not guaranteed that the error due to the factorization approximation is always of higher order than the order of our interest.

\section{Poisson brackets of the standard variables}\label{app:standard variables}
If we use the canonical pairs $c^I:=\mean{\hat{c}^I}$
and $p_I:=\mean{\hat{p}_I}$ as the classical variables,
the associated quantum variables are defined as:
\ba
G^{(n=2)}:\qquad\quad
G_{IJ}&:=&\mean{(\hat{p}_I-\mean{\hat{p}_I})(\hat{p}_J-\mean{\hat{p}_J})},\nn\\
G^{IJ}&:=&\mean{(\hat{c}^I-\mean{\hat{c}^I})(\hat{c}^J-\mean{\hat{c}^J})},\nn\\
G^I_J&:=&\mean{(\hat{c}^I-\mean{\hat{c}^I})(\hat{p}_J-\mean{\hat{p}_J})}_\mathrm{Weyl}.
\ea

This is the standard formalism, in which the quantum variables commute with the classical variables. For reference, the Poisson brackets of the standard quantum variables of the 2nd order are listed:
\ba
&&\{G^I_J,G^K_L\}=(8\pi G\gamma)^2\left(\delta^I_L G^K_J-\delta^K_J G^I_L\right),
\quad
\{G^{IJ},G_{KL}\}=4 (8\pi G\gamma)^2 \delta^{(I}_{(K} G^{J)}_{L)},\nn\\
&&\{G^{IJ},G^K_L\}=2 (8\pi G\gamma)^2 \delta^{(I}_{L} G^{J)K},\qquad
\{G_{IJ},G^K_L\}=-2(8\pi G\gamma)^2 \delta^{K}_{(I} G_{J)L},\nn\\
&&\{G_{IJ},G_{KL}\}=\{G^{IJ},G^{KL}\}=0.
\ea

\end{document}